%% file: main.tex
\documentclass[5p,times]{elsarticle}

\usepackage[ruled,vlined]{algorithm2e}
\usepackage{amsmath} 
\usepackage{amssymb} 
\usepackage[
  final,
  tracking=true,
  expansion=true,
  protrusion=true
]{microtype}
\usepackage[utf8]{inputenc}
\usepackage{amsmath, amssymb, amsfonts}
\usepackage{graphicx}
\usepackage{booktabs} 
\usepackage{multirow}
\usepackage[most]{tcolorbox}
\usepackage{dblfloatfix}
\usepackage{tabularx} 
\usepackage{amsmath}
\usepackage{amssymb}
\usepackage{amsfonts}
\usepackage{multirow}
\usepackage{longtable}
\usepackage{caption}
\usepackage{subcaption}
\usepackage{pifont}
\usepackage{array}
\usepackage{url}
\usepackage{xcolor}
\usepackage[colorlinks=true,allcolors=white]{hyperref}
\usepackage{makecell}

\newcolumntype{P}[1]{>{\centering\arraybackslash}p{#1}}

\usepackage[switch]{lineno}

\usepackage{booktabs}   
\usepackage{multirow}   
\usepackage{graphicx}   
\usepackage{array}      
\usepackage{xcolor}     
\usepackage{balance}

\begin{document}

\let\WriteBookmarks\relax
\def\floatpagepagefraction{1}
\def\textpagefraction{.001}


\title{Red-MIRROR: Agentic LLM-based Autonomous Penetration Testing with Reflective Verification and Knowledge-augmented Interaction}                      



%

\affiliation[1]{organization={Information Security Lab, University of Information Technology},
    city={Ho Chi Minh City},
    country={Vietnam}}
\affiliation[2]{organization={Vietnam National University},
    city={Ho Chi Minh City},
    country={Vietnam}}

\author[1,2]{Tran Vy Khang}
\ead{22520628@gm.uit.edu.vn}

\author[1,2]{Nguyen Dang Nguyen Khang}
\ead{22520617@gm.uit.edu.vn}

\author[1,2]{Nghi Hoang Khoa}
\ead{khoanh@uit.edu.vn}

\author[1,2]{Do Thi Thu Hien}
\ead{hiendtt@uit.edu.vn}

\author[1,2]{Van-Hau Pham}
\ead{haupv@uit.edu.vn}

\author[1,2]{Phan The Duy\corref{cor1}}
\ead{duypt@uit.edu.vn}
\cortext[cor1]{Corresponding author}

\begin{abstract}


Web applications remain the dominant attack surface in cybersecurity, where vulnerabilities such as SQL injection, XSS, and business logic flaws continue to cause significant data breaches. While penetration testing is effective for identifying these weaknesses, traditional manual approaches are time-consuming and heavily dependent on scarce expert knowledge. Recent Large Language Models (LLM)-based multi-agent systems have shown promise in automating penetration testing, yet they still suffer from critical limitations: over-reliance on parametric knowledge, fragmented session memory, and insufficient validation of attack payloads and responses.
This paper proposes Red-MIRROR, a novel multi-agent automated penetration testing system that introduces a tightly coupled memory–reflection backbone to explicitly govern inter-agent reasoning. By synthesizing Retrieval-Augmented Generation (RAG) for external knowledge augmentation, a Shared Recurrent Memory Mechanism (SRMM) for persistent state management, and a Dual-Phase Reflection mechanism for adaptive validation, Red-MIRROR provides a robust solution for complex web exploitation. Empirical evaluation on the XBOW benchmark and Vulhub CVEs shows that Red-MIRROR achieves performance comparable to state-of-the-art agents on Vulhub scenarios, while demonstrating a clear advantage on the XBOW benchmark. On the XBOW benchmark, Red-MIRROR attains an overall success rate of 86.0 percent, outperforming PentestAgent (50.0 percent), AutoPT (46.0 percent), and the VulnBot baseline (6.0 percent). Furthermore, the system achieves a 93.99 percent subtask completion rate, indicating strong long-horizon reasoning and payload refinement capability. Finally, we discuss ethical implications and propose safeguards to mitigate misuse risks.


\end{abstract}




\maketitle

\input{sections/1-introduction}
\input{sections/2-background}
\input{sections/3-methodology}

\input{sections/4-experiment}
\input{sections/5-discussion}
\input{sections/6-conclusion}








\bibliographystyle{unsrt}

\bibliography{refs.bib}
\balance

\end{document}

%% file: sections/1-introduction.tex
\section{Introduction} \label{sec_introduction}
The rapid proliferation of web applications has substantially expanded the externally accessible attack surface of modern digital infrastructures. The Verizon Data Breach Investigations Report (DBIR) 2025 identifies Basic Web Application Attacks as one of the five most prevalent breach patterns, accounting for approximately 12\% of over 12,000 confirmed data breaches analyzed worldwide \cite{verizon2025dbir}. In absolute terms, this corresponds to more than one thousand real-world compromise incidents attributable to web-facing systems, demonstrating that web applications remain a consistently exploited entry point in large-scale breach scenarios. Trends in vulnerability reporting further highlight the prominence of application-layer weaknesses. According to OWASP’s Top 10 classifications, the number of publicly recorded CVE entries mapped to Injection-related weaknesses increased from approximately 32,000 in 2021 to around 64,000 in 2025, reflecting a substantially larger corpus of publicly documented injection vulnerabilities in recent vulnerability datasets \cite{owasp2021top10,owasp2025top10}. Given their inherent public accessibility and direct exposure to untrusted inputs, vulnerabilities such as SQL Injection (SQLi) and Cross-Site Scripting (XSS) continue to present a critical security challenge, enabling attackers to compromise sensitive user and enterprise data \cite{owasp2025top10,owasp2023wstg}.

Penetration testing is widely regarded as one of the most effective approaches for identifying web security vulnerabilities, assessing their severity, and proposing appropriate mitigation strategies \cite{altulaihan2023survey}. However, traditional penetration testing often requires organizations to invest significant time and financial resources in building and maintaining highly skilled security teams with years of practical experience \cite{isc22025workforce, zhu2025software}.

To address these limitations in cost and human resources, automated security testing solutions have been actively studied and developed, particularly those integrating Large Language Models (LLMs) with multi-agent systems \cite{guo2024multiagents, deng2024pentestgpt, zhou2025large}. Such LLM-based approaches demonstrate strong capabilities in penetration testing through effective pattern-matching for vulnerability detection, handling uncertainty in dynamic environments, and cost-effective integration \cite{happe2025surprising}. While the automation of penetration testing offers significant promise for defensive security, it also introduces dual-use risks, as the same capabilities could be repurposed for malicious activities \cite{sheng2025llms}. Consequently, the development of such systems must be framed within a ``Responsible AI Research'' paradigm, emphasizing their role as tools for red teaming, vulnerability discovery, and proactive defense rather than enabling cybercrime. Recent studies such as VulnBot \cite{kong2025vulnbot}, PentestAgent \cite{shen2025pentestagent}, xOffense \cite{luong2025xoffense}, Autopentest \cite{benazzouz2025autopentest}, AutoPT \cite{Wu2025AutoPT} and PTFusion \cite{wang2025ptfusion} demonstrate that LLM-based agents, when combined with multi-agent architectures and external knowledge sources through Retrieval-Augmented Generation (RAG), are capable of effectively exploiting web application vulnerabilities. 

However, despite the above advancements, several fundamental architectural bottlenecks hinder their reliability and effectiveness in real-world scenarios, which can be categorized into three critical limitations. First, existing systems suffer from inefficient memory and session management \cite{kong2025vulnbot, shen2025pentestagent, Wu2025AutoPT}. Information maintained within a single conversational history often leads to memory fragmentation, resulting in degraded performance in long-horizon attack scenarios. As interactions accumulate, critical tokens from early reconnaissance phases may be truncated or diluted by later outputs, preventing the agent from effectively linking initial discoveries to subsequent exploitation steps. For example, VulnBot \cite{kong2025vulnbot} reports challenges related to context loss and session continuity, particularly in multi-phase penetration workflows. This limitation hampers coherent long-horizon reasoning \cite{liu2023lost}, which is essential for executing complex, multi-step attack chains. Second, current approaches lack robust quality control for request handling \cite{kong2025vulnbot, shen2025pentestagent, Wu2025AutoPT}. Without a mechanism to validate payloads before execution, agents can repeatedly issue malformed requests, increasing detection risks. For instance, selecting an invalid injection point can prematurely fail an entire testing workflow without meaningful feedback. Third, integrated testing tools are often insufficiently specialized for modern applications. Automated frameworks typically rely on general-purpose command-line tools such as Nmap or SQLMap, which lack the granularity needed to test complex, dynamic application logic found in contemporary web environments.

To bridge this gap and address these critical architectural limitations, we propose \textbf{Red-MIRROR} (\textbf{Red}-teaming \textbf{M}ulti-agent \textbf{I}ntrospective \textbf{R}easoning for \textbf{R}obust \textbf{O}ffensive \textbf{R}esearch), an automated multi-agent penetration testing system that enhances memory and session management during inter-agent interactions and testing workflows. In particular, our framework leverages Retrieval-Augmented Generation (RAG) to provide external knowledge relevant to penetration testing.
In addition, Red-MIRROR introduces mechanisms for validating, analyzing, and refining requests and attack plans based on observed responses during testing, thereby establishing a closed-loop feedback process. 
Finally, Red-MIRROR integrates a set of specialized web testing tools, further improving its effectiveness in exploiting vulnerabilities in web applications.


Moreover, we contribute by constructing a fine-tuning and training dataset for a mid-scale LLM to evaluate the potential of self-hosted open-source models in comparison with commercial models. We also develop a subtask-level benchmark based on a subset of the XBOW \cite{xbow2024} benchmark and selected CVEs from Vulhub \cite{vulhub2025}, allowing for a more detailed evaluation of test coverage and step-by-step execution capabilities across the penetration testing workflow.

In summary, we make the following four key contributions:

\begin{itemize}
    \item We introduce \textbf{Red-MIRROR}, the first multi-agent penetration testing framework that tightly couples a Shared Recurrent Memory Mechanism (SRMM) with Dual-Phase Reflection and RAG-augmented knowledge. This backbone explicitly governs inter-agent reasoning and adaptive validation, overcoming memory fragmentation and payload hallucination in long-horizon attacks. Extensive evaluation on the XBOW benchmark demonstrates an overall exploitation success rate of 86.0\% (vs. 50.0\% PentestAgent, 46.0\% AutoPT, and 6.0\% VulnBot baseline) together with a 93.99\% subtask completion rate, establishing a new state-of-the-art in autonomous web penetration testing.

    \item We curate a specialized, high-fidelity fine-tuning dataset of 1,644 prompt-response pairs covering CVE descriptions, CAPEC attack patterns, and MITRE ATT\&CK techniques. By applying LoRA on Qwen2.5-14B, we demonstrate that a mid-scale open-source LLM can achieve competitive pentesting performance against commercial models, significantly lowering the barrier for self-hosted offensive AI research.

    \item We construct a fine-grained subtask-level benchmark derived from XBOW subsets and real-world Vulhub CVEs. This benchmark enables precise measurement of reconnaissance, exploitation, and reflection capabilities across multi-step workflows, addressing a critical evaluation gap in prior LLM-based pentesting studies.

    \item We explicitly address the dual-use risks of offensive AI through a comprehensive ethical analysis and propose practical safeguards (role-based access control, audit logging, and RAG knowledge gating) that ensure responsible deployment of Red-MIRROR in authorized red-teaming environments.
\end{itemize}




The remainder of this paper is organized as follows. \textbf{Section~\ref{sec:background}} reviews the background and related work on automated penetration testing and LLM-based agents. \textbf{Section~\ref{sec:methodology}} describes the architecture of our Red-MIRROR framework, detailing the components for memory management and reflective reasoning. The experimental results and comparative analysis are presented in \textbf{Section~\ref{sec:experiments}}. \textbf{Section~\ref{sec:discussion}} provides a discussion on the findings, limitations, and ethical implications. Finally, we conclude our research and suggest future directions in \textbf{Section~\ref{sec:conclusion}}.

%% file: sections/2-background.tex
\section{Background and Related Work}
\label{sec:background}

\subsection{Penetration Testing and Automation Challenges}
\label{sec:background_pentest}

Penetration testing (pentest) is a comprehensive security assessment methodology that evaluates the robustness of computer systems, networks, and applications by simulating real-world adversarial attacks. Unlike automated vulnerability scanning, which primarily detects known weaknesses using signature-based techniques, penetration testing emphasizes exploitability, impact assessment, and contextual risk validation through adaptive attack execution \cite{nist800115}. This process often requires human expertise to interpret system behavior, adjust strategies, and reason over multiple intermediate findings.

Standardized frameworks such as the Penetration Testing Execution Standard (PTES) \cite{ptes2014} and NIST guidelines define penetration testing as a multi-phase and iterative process, typically encompassing reconnaissance, scanning and analysis, exploitation, and reporting. These phases are tightly interdependent: information gathered during exploitation frequently informs further reconnaissance, while partial failures may require reformulating attack strategies. In web application security, effective penetration testing additionally demands session awareness, understanding of application logic, and dynamic payload adaptation in response to server-side behaviors. Vulnerabilities such as SQL Injection, Cross-Site Scripting (XSS), and authentication bypasses often require multi-step interactions rather than single-shot exploits.

Despite its effectiveness, traditional penetration testing is time-consuming, labor-intensive, and difficult to scale across large or continuously evolving systems. Modern web applications frequently undergo rapid development cycles, making manual testing insufficient to keep pace with emerging vulnerabilities. As a result, there is a growing demand for automated pentesting solutions that can replicate expert reasoning while maintaining efficiency and scalability.  Recent studies such as VulnBot \cite{kong2025vulnbot}, PentestAgent \cite{shen2025pentestagent}, xOffense \cite{luong2025xoffense}, Autopentest \cite{benazzouz2025autopentest}, AutoPT \cite{Wu2025AutoPT} and PTFusion \cite{wang2025ptfusion} demonstrate that LLM-based agents, when combined with multi-agent architectures and external knowledge sources through Retrieval-Augmented Generation (RAG), can effectively exploit web application vulnerabilities. 

However, despite these advancements, these systems exhibit several common limitations. First, systems such as VulnBot, PentestAgent, and xOffense suffer from inefficient memory and session management still leads to context fragmentation in long-horizon multi-phase workflows. Second, approaches including VulnBot, AutoPT, and Autopentest lack robust validation mechanisms, often resulting in blind or malformed payload execution without systematic verification due to persistent LLM hallucinations and insufficient auto-checks. Third, existing frameworks such as AutoPT, and PentestAgent rely on loosely integrated or general-purpose tools leading to inefficient and inconsistent exploitation behavior caused by imprecise command generation and heterogeneous output fusion.

Automating penetration testing remains challenging due to its reliance on long-horizon reasoning, uncertainty management, and expert decision-making \cite{liu2023lost, valmeekam2023planning}. While recent systems have incorporated scripted workflows or rule-based engines, such approaches struggle to generalize across heterogeneous targets and complex application logic. More recent research has explored the use of intelligent agents to replicate portions of expert reasoning, enabling automated systems to conduct reconnaissance, select candidate attack vectors, and validate exploitation outcomes. However, existing automated pentesting frameworks often suffer from fragmented context management, limited adaptability, and unreliable verification of exploit success, particularly in extended testing sessions. These limitations motivate the exploration of reasoning-centric and memory-aware approaches that can better capture the iterative and stateful nature of real-world penetration testing.

\subsection{Large Language Models for Security Reasoning and Knowledge Augmentation}
\label{sec:background_llm}

LLMs are neural architectures trained on large-scale textual corpora to perform a wide range of language understanding and generation tasks. Most contemporary LLMs are built upon the Transformer architecture, which employs self-attention mechanisms to model long-range dependencies within input sequences \cite{vaswani2017attention}. These properties make LLMs suitable for security-related applications, including vulnerability reasoning, payload generation, HTTP request analysis, and interpretation of server responses \cite{silva2024survey}.

Two primary strategies are commonly used to adapt LLMs to domain-specific tasks: prompt engineering and fine-tuning. Prompt engineering exploits in-context learning by carefully constructing input prompts that include structured instructions or few-shot examples, guiding model behavior without modifying model parameters. While this approach enables rapid experimentation and deployment, it is constrained by finite context windows and often exhibits unstable performance when applied to complex, multi-step reasoning tasks such as penetration testing workflows \cite{liu2023lost}.

Fine-tuning, in contrast, involves further training a pre-trained LLM on task-specific datasets to align its internal representations with domain requirements. Parameter-efficient fine-tuning techniques, such as Low-Rank Adaptation (LoRA) \cite{hu2021lora}, significantly reduce computational costs while enabling effective specialization. In security contexts, fine-tuning can enhance a model’s understanding of exploit patterns, vulnerability taxonomies, and structured output formats, leading to improved consistency and reliability in automated systems.

Despite these advances, LLMs inherently suffer from knowledge staleness and hallucination, particularly in fast-evolving domains such as cybersecurity. Retrieval-Augmented Generation (RAG) \cite{lewis2020rag} addresses these limitations by integrating external knowledge retrieval mechanisms into the generation process. By augmenting prompts with relevant documents retrieved from curated knowledge bases, RAG enables LLMs to access up-to-date vulnerability descriptions, exploit techniques, and testing guidelines without frequent retraining. In penetration testing applications, grounding model outputs in authoritative sources such as OWASP documentation or CVE databases improves factual accuracy, reduces hallucinations, and enables smaller or mid-scale LLMs to achieve competitive performance in specialized security tasks.

\subsection{Multi-Agent Architectures with Memory and Reflection Mechanisms}
\label{sec:background_multiagent}

Multi-agent systems consist of multiple autonomous agents that collaborate to achieve complex objectives through task decomposition and coordination \cite{guo2024multiagents}. This paradigm closely mirrors real-world penetration testing practices, where reconnaissance, exploitation, and analysis are typically performed by specialists with distinct responsibilities. Recent LLM-based pentesting frameworks increasingly adopt multi-agent architectures to distribute roles such as attack planning, payload execution, and response interpretation, improving modularity and scalability.

However, naive multi-agent implementations often rely on conversational message passing as the primary coordination mechanism. Such designs can lead to redundant communication, inconsistent system states, and loss of critical contextual information during long-horizon tasks \cite{xi2023rise}. These issues are particularly pronounced in penetration testing, where maintaining session state, tracking discovered endpoints, and reasoning over prior exploit attempts are essential for effective attack planning.

To address these challenges, recent research has explored memory-augmented and recurrent Transformer architectures that introduce shared or persistent memory abstractions for long-horizon reasoning~\cite{sagirova2025srmt}. Shared Recurrent Memory Mechanism (SRMM)–inspired approaches enable agents to accumulate salient information over time and access a common contextual state without relying solely on explicit message exchanges. At a system level, such shared memory mechanisms support consistent session handling, reduce redundant exploration, and enhance coordination across agents during extended penetration testing workflows.

In addition to memory, reflection mechanisms have been proposed to enable intelligent agents to evaluate outcomes and refine future actions based on feedback~\cite{guo2025mirror}. Reflection is particularly important in automated penetration testing to avoid repeated failures, validate exploit attempts, and adapt strategies to dynamic defenses such as input sanitization or Web Application Firewalls. Dual-phase reflection paradigms conceptually separate local, short-term evaluation from global, long-term strategy refinement. By combining intra-reflection, which assesses the correctness of immediate actions, with inter-reflection, which analyzes outcomes across agents and iterations, automated systems can iteratively improve decision-making quality. Together, memory-centric coordination and reflection-based adaptation form a conceptual foundation for more reliable, stable, and scalable automated penetration testing systems. In contrast to prior systems that primarily introduce reflection as a local heuristic or prompt-level correction~\cite{shen2025pentestagent}, Red-MIRROR operationalizes reflection as a system-level control mechanism tightly coupled with persistent shared memory and planning.

%% file: sections/3-methodology.tex
\section{Methodology} \label{sec:methodology}


\subsection{Motivating Example: Long-Context Chained Exploitation}
\label{sec:motivating_example}
To illustrate the challenges of long-horizon reasoning in autonomous penetration testing, consider a benchmark instance involving a reflected XSS vulnerability protected by server-side filtering.

During the reconnaissance phase, the agent accesses the main page and discovers the endpoint \texttt{/page}. A request to \texttt{/page?name=test} reveals that the parameter \texttt{name} is reflected inside an HTML attribute as follows, confirming that user input is embedded within a quoted attribute context.

\begin{center}
\texttt{<input type=text name=name placeholder="Enter your name..." value="test">}
\end{center}

The exploitation phase requires iterative experimentation. Initial payloads such as
\texttt{<script>alert('XSS')</script>} are sanitized by the server. Subsequent attempts reveal a systematic transformation: any injected substring matching the pattern \texttt{<[a-z/]} is removed or neutralized before rendering. For example, inputs beginning with \texttt{<script}, \texttt{<img}, or \texttt{</input>} consistently disappear or are replaced with empty strings in the reflected output. Across approximately 20–30 consecutive attempts, the agent observes that:

\begin{itemize}
    \item The leading character \texttt{<} followed by an alphabetical character or slash is consistently stripped.
    \item Attribute-breaking characters such as quotation marks remain partially intact.
    \item Non-tag-based characters (like alphanumeric strings or certain symbols) are preserved.

\end{itemize}

This repeated behavior allows the agent to infer that the server employs a pattern-based filter targeting HTML tag openings rather than performing full contextual encoding.

In Red-MIRROR, these execution outcomes are not treated as isolated conversational turns. Through persistent structured memory and response-driven reflection, the system accumulates transformation evidence across attempts, gradually refining its internal hypothesis about the filtering rule. Instead of repeatedly generating semantically equivalent tag-based payloads, the agent learns to avoid direct \texttt{<tag>} constructs and instead focuses on escaping the attribute boundary using preserved characters (like quotation marks) before injecting an event-triggered handler that does not match the filtered prefix pattern. When rendered, the evaluation environment automatically activates the injected handler, resulting in successful JavaScript execution and flag retrieval.

\begin{figure*}[!b]
\centering
\includegraphics[width=1\textwidth]{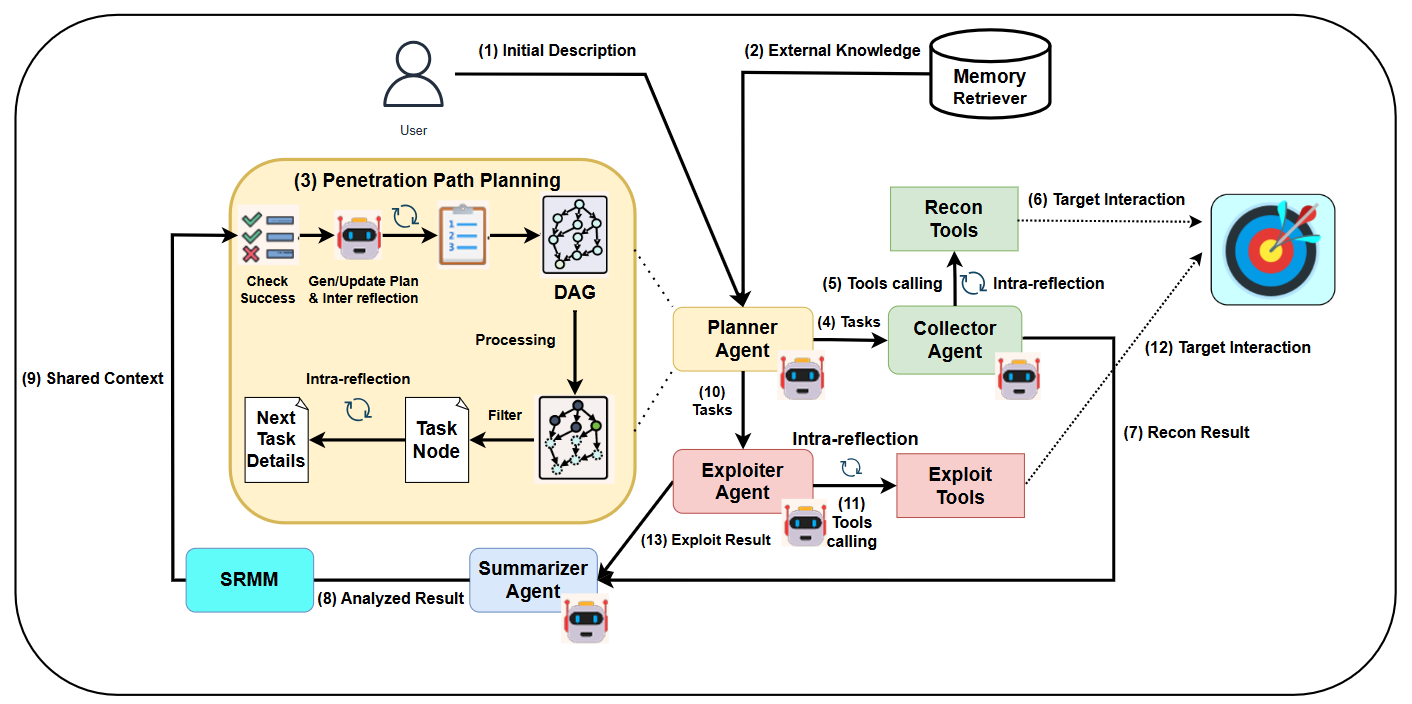}
\caption{The holistic architecture of the Red-MIRROR framework. The system is centered around a shared reasoning backbone that integrates persistent memory and dual-phase reflection, upon which planning and execution agents operate within a cyclic decision-making workflow.}
\label{fig:architecture}
\end{figure*}

However, in systems relying solely on standard conversational histories without specialized state management, such cumulative reasoning is fragile. After dozens of interaction turns—including failed payloads, sanitized responses, and routine HTTP outputs—earlier observations become diluted within an expanding context window. Without explicit consolidation of prior transformation patterns, agents may inadvertently regenerate inputs beginning with \texttt{<[a-z/]} or revisit previously invalidated tag-based strategies. Instead of progressively narrowing the hypothesis space, exploration becomes cyclical, with repeated failures and diminished convergence probability.

This example demonstrates that successful chained exploitation in filtered environments depends not only on generative capability, but on sustained state retention and iterative reasoning over extended horizons. It motivates the need for persistent shared memory and structured reflection mechanisms in autonomous pentesting systems, forming the foundation of Red-MIRROR's architectural design.



\subsection{Threat Model}
\label{sec:threat_model}
Building upon the motivating example's illustration of long-horizon challenges in exploiting reflected XSS under server-side filtering (such as pattern-based sanitization leading to memory dilution and cyclical failures), we formalize the assumptions, boundaries, and risks in Red-MIRROR using the STRIDE threat modeling framework \cite{kohnfelder1999threats, hernan2006stride}. STRIDE categorizes threats into six classes: Spoofing (impersonation), Tampering (data alteration), Repudiation (denial of actions), Information Disclosure (unauthorized leaks), Denial of Service (DoS, resource exhaustion), and Elevation of Privilege (EoP, unauthorized escalation). This model is well-suited for AI-driven pentest systems, addressing both simulated attacks on web applications (such as SQLi/XSS from OWASP Top 10, as in Section 2) and architectural risks (like LLM hallucinations or context fragmentation in multi-agent workflows).

We assume a black-box ethical red-teaming environment, where the target is a web application with vulnerabilities accessible via HTTP/HTTPS, and the simulated adversary (Red-MIRROR) uses public tools (such as Nmap, SQLMap) without insider privileges. \label{sec:pentest_objectives} Uncertainties include dynamic defenses like input sanitization or Web Application Firewalls (WAFs, such as ModSecurity), as exemplified by the motivating example's filtering patterns—though not fully tested in our benchmarks (a limitation for future work, aligning with PTES guidelines in Section \ref{sec:background_pentest}). The system emphasizes defensive deployment with user consent to mitigate dual-use risks.

To systematically address dual-use risks inherent in agentic pentest systems (such as memory tampering leading to cyclical failures, or reflection loops causing DoS), we adopt a lightweight STRIDE-inspired analysis. Key threats are mitigated as follows: (i) Tampering \& Information Disclosure countered by SRMM’s immutable write-once semantic and gated RAG; (ii) Denial-of-Service prevented by Dual-Phase Reflection’s hard termination (max 10 iterations); (iii) Elevation-of-Privilege restricted via non-destructive mode and post-execution verification in Exploiter Agent; (iv) Spoofing \& Repudiation are considered out-of-scope under our black-box threat model, as the system does not involve identity-level authentication or user-attribution mechanisms.


\subsection{Overall System Architecture} \label{subsec:overall_architecture}

\subsubsection{Architecture Overview}
The proposed Red-MIRROR (\textbf{Red}-teaming \textbf{M}ulti-agent \textbf{I}ntrospective \textbf{R}easoning for \textbf{R}obust \textbf{O}ffensive \textbf{R}esearch) framework, as in \textbf{Fig.~\ref{fig:architecture}}, is designed as a stateful, cyclic multi-agent system for automated penetration testing, implemented on top of the LangGraph orchestration library \cite{langgraph2025}. In contrast to conventional linear or stage-based execution pipelines, Red-MIRROR adopts a graph-oriented computational flow that explicitly supports iteration, feedback, and strategic revision \cite{besta2024graph}.

The core of Red-MIRROR’s operational effectiveness lies in the tight coupling of two key components: a Shared Recurrent Memory Mechanism (SRMM) and a Dual-phase reflection mechanism. The SRMM maintains a unified global state that persistently aggregates reconnaissance findings, intermediate reasoning artifacts, and exploitation outcomes across the entire engagement. Built upon this persistent memory, the Dual-phase reflection mechanism governs the system’s reasoning dynamics, continuously refining strategy through planner-level introspection.

The workflow initializes from a task-specific objective, transitions into an iterative reconnaissance phase guided by planning–execution loops, and culminates in targeted exploitation attempts. To further ground reasoning and reduce hallucinated decisions, Red-MIRROR integrates a Retrieval-Augmented Generation (RAG) module \cite{lewis2020rag} that incorporates external vulnerability intelligence. While planning, execution, and knowledge retrieval are handled by specialized agents, their behavior is coordinated by the shared memory and Dual-phase reflection mechanisms that form the conceptual core of the framework.

\subsubsection{Planner Agent}
The Planner Agent acts as the strategic reasoning entity within Red-MIRROR, responsible for translating high-level objectives into structured, executable tasks while continuously adapting to newly acquired evidence. Rather than generating a static end-to-end attack plan, the Planner incrementally constructs and revises its strategy based on the evolving global state maintained in the SRMM.

A central function of the Planner is the management of the Penetration Path Planning module, which represents the attack process as a dynamically evolving directed acyclic graph (DAG) \cite{yao2024tree}. In this graph, each node corresponds to a context-aware task, such as focused service enumeration or vulnerability validation, while edges encode logical and operational dependencies. This explicit structure enables the Planner to reason about prerequisite satisfaction, execution order, and alternative paths when a given strategy proves ineffective.

In addition, the Planner performs inter-agent reflection by analyzing outputs produced by downstream agents. When discrepancies arise between expected and observed outcomes, the Planner revises the attack graph by pruning unproductive branches, introducing alternative reconnaissance actions, or querying external knowledge sources. This reflective capability ensures that Red-MIRROR remains adaptive, goal-driven, and resilient to uncertainty throughout the engagement lifecycle.

\subsubsection{Collector Agent}
The Collector Agent is responsible for discovering, enumerating, and characterizing the target’s attack surface. To minimize fragmentation and context loss, Red-MIRROR consolidates reconnaissance and scanning responsibilities within this single agent, reducing inter-agent communication overhead and preserving temporal coherence between discovery and exploitation.

Guided by planner-issued tasks, the Collector executes narrowly scoped reconnaissance actions using established tools such as network scanners (like Nmap) and web enumeration utilities (like Dirsearch). Rather than performing exhaustive scans, each action is explicitly motivated by the current penetration path, reducing noise and limiting unnecessary exposure to defensive mechanisms.

The Collector normalizes raw tool outputs into structured observations, including identified services, version fingerprints, exposed endpoints, and configuration anomalies. These observations are committed to the SRMM, forming a reliable empirical foundation for subsequent planning and exploitation decisions. The agent also infers higher-level contextual attributes, such as technology stacks and deployment patterns, which are critical for accurate vulnerability assessment.

\subsubsection{Exploiter Agent} 
The Exploiter Agent represents the active offensive capability of the Red-MIRROR system. Its primary objective is to generate, validate, and execute payloads tailored to the vulnerabilities identified during the reconnaissance phase.

The defining characteristic of the Exploiter is its Intra-reflection capability. Before a payload is dispatched to the target, the agent performs a self-assessment to ensure the payload's syntax and logic align with the identified vulnerability context. Post-execution, the agent meticulously analyzes the target's response such as HTTP status codes, timing differences, or specific error strings. If an attempt is unsuccessful, the Exploiter uses this feedback to mutate the payload autonomously—applying different encoding schemes or bypass techniques—thereby increasing the robustness of the attack while minimizing the number of invalid requests.

\subsubsection{Summarizer (Analyzer) Agent} The Summarizer (Analyzer) Agent which is integrated within the reflection and reporting flow serves as the system's “Knowledge Synthesis” layer. Its role is to distill the vast amount of raw data and execution logs stored in the SRMM into actionable intelligence and final reports.

During the transition between reconnaissance and exploitation, this agent performs a critical Knowledge Transfer Analysis. It correlates disparate findings, such as a specific service version from the Collector and a known exploit from the RAG module—to provide the Planner with a prioritized list of attack vectors. At the conclusion of a session, the Summarizer generates a comprehensive report that outlines the successful exploitation paths, the vulnerabilities discovered, and the reflection steps taken, providing a clear audit trail of the system's reasoning process.

\subsection{Shared Recurrent Memory Mechanism}
\label{subsec:srmm}

\subsubsection{Overview}

To ensure coherent coordination across reconnaissance and exploitation phases while mitigating context fragmentation, MIRROR employs a Shared Recurrent Memory Mechanism (SRMM). Inspired by the Shared Recurrent Memory Transformer (SRMT)~\cite{sagirova2025srmt}, SRMM is adapted as a lightweight, text-based shared state optimized for LLM-based multi-agent systems under constrained resources. Unlike conventional shared logs or conversation histories, SRMM enforces explicit provenance-preserving partitioning, unidirectional information flow, and bounded aggregation—properties that prevent redundancy, feedback loops, and unbounded context growth.

\begin{figure}[!b]
    \renewcommand{\figurename}{Figure}
    \centering
    \includegraphics[width=1\linewidth]{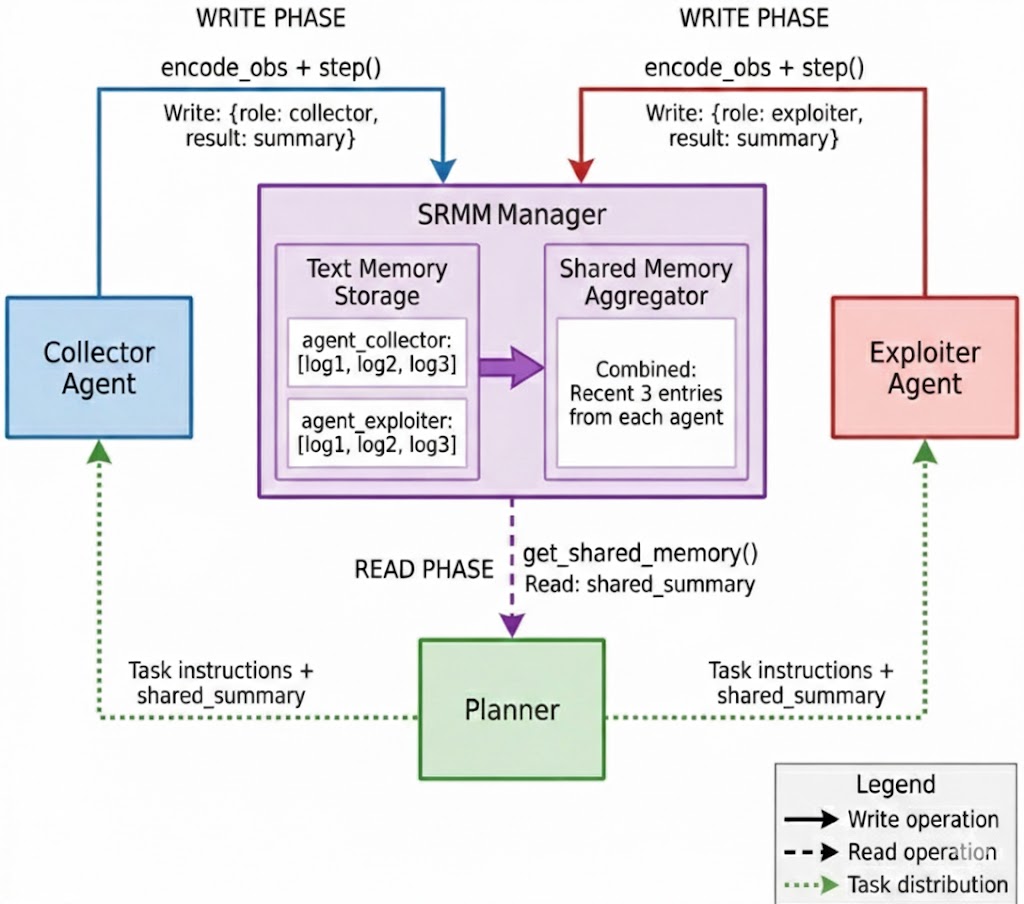}
    \caption{The data flow of SRMM in Red-MIRROR system.}
    \label{fig:srmm_flow}
\end{figure}

As shown in \textbf{Fig.~\ref{fig:srmm_flow}}, SRMM comprises three core components of (1) \textit{Text Memory Storage} which is the partitioned history, (2) \textit{Shared Memory Aggregator} - a summarization function $\Sigma$ producing compact global context, and (3) \textit{SRMM Manager} enforcing access control and unidirectional semantics. Moreover, from the view of this mechanism, agents are partitioned into execution agents $\mathcal{A}_E$ and planner agents. While the former, including Collector and Exploiter, aim to perform specific actions, the latter is responsible for making plans. The mechanism is also designed with two phases, with the awareness of the above partition. In particular, the Write phase involves only execution agents $\mathcal{A}_E$ appending observations, while the planner agent may query the aggregated context in the Read phase.

Formally, SRMM maintains a shared \textit{Text Memory Storage} $\mathcal{M}_{\text{text}}$ as an agent-partitioned, temporally indexed history as in \eqref{eq:srmm-memory}.
\begin{equation}
\mathcal{M}_{\text{text}} = \bigcup_{a \in \mathcal{A}_E} \left\{ (a, t, h_{a,t}) \;\middle|\; t \in \mathbb{N}_0,\ h_{a,t} \in \mathcal{H}_a \right\}
\label{eq:srmm-memory}
\end{equation}
where $\mathcal{A}_{E}$ is the set of execution agents, $\mathcal{H}_a$ is the observation language of agent $a$ (structured text/JSON), and $t$ indexes local writes per agent.



\subsubsection{Operators, and Formal Properties}

There are two types of operators available in the designed SRMM, including write and read operators. 
\begin{itemize}
    \item \textit{Write operator}  is defined as a monotonic append-only function, as in \eqref{eq:write-op}, where $o$ is serialized into $s \in \mathcal{H}_a$ and appended at the current timestep $t_{\text{now}}$ (local to agent $a$) (\eqref{eq:time_append}). This operator is concretely implemented in \textbf{Algorithm~\ref{alg:srmm_write}}, which handles serialization fallback and atomic appends to maintain provenance.
\begin{equation}
\mathcal{W}(a, o) : \mathcal{M}_{\text{text}} \to \mathcal{M}_{\text{text}}',\quad a \in \mathcal{A}_E
\label{eq:write-op}
\end{equation}
\begin{equation}
\mathcal{M}_{\text{text}}' = \mathcal{M}_{\text{text}} \cup \bigl\{(a, t_{\text{now}}, s)\bigr\},\quad h_{a,t_{\text{now}}} = s.
\label{eq:time_append}
\end{equation}

\begin{algorithm}[!t]
\LinesNotNumbered
\caption{Write Observation to SRMM}
\label{alg:srmm_write}
\SetKwInOut{KwInput}{Input}
\SetKwInOut{KwOutput}{Output}

\textbf{function} \texttt{WriteObservation}($a$, $o$)

\KwInput{Agent $a \in \mathcal{A}_E$, observation $o$}
\KwOutput{Updated $\mathcal{M}_{\text{text}}$}

\eIf{successful JSON serialization}{
    $s \gets \texttt{json.dumps}(o)$\;
}{
    $s \gets \texttt{str}(o)$\;
}

\If{$a \notin \mathcal{M}_{\text{text}}$}{
    $\mathcal{M}_{\text{text}}[a] \gets []$\;
}

$\mathcal{M}_{\text{text}}[a].\texttt{append}(s)$\;

\KwRet \texttt{GetSharedMemory}()\;
\end{algorithm}
\item \textit{Read operator} is defined as a read-only retrieval and aggregation function accessible only to the planner $p$ as in \eqref{eq:read-op}, where $\mathcal{S}_{\text{summary}} \subseteq \mathcal{H}_p^*$ denotes the space of summarized planner contexts (natural-language strings).
\begin{equation}
\mathcal{R}(p, k) : \{p\} \times \mathbb{N} \to \mathcal{S}_{\text{summary}},
\label{eq:read-op}
\end{equation}
  
It first applies the deterministic filter to extract the $k$ most recent entries per agent (based on each agent's local timeline), then formats and aggregates them before summarization as in \eqref{eq:summary}.
\begin{equation}
c_{\text{summary}} = \Sigma \circ \text{Format} \circ \text{Filter}_k (\mathcal{M}_{\text{text}}).
\label{eq:summary}
\end{equation}
Here, $\text{Filter}_k(\mathcal{M}_{\text{text}}) = \bigcup_{a \in \mathcal{A}_E} \{ h_{a,t} \mid t \geq T_a - k + 1 \}$,  
with $T_a = \max\{ t \mid (a, t, h_{a,t}) \in \mathcal{M}_{\text{text}} \}$ denoting the latest local timestep for agent $a$. The $\text{Format}$ step tags each observation with its agent prefix 
and concatenates them into a single string.

The read operator is side-effect free when $\mathcal{R}(p,k)$ does not modify $\mathcal{M}_{\text{text}}$. It is concretely realized in \textbf{Algorithm~\ref{alg:srmm_read}}, which returns a combined string ready for LLM summarization.
\end{itemize}
\begin{algorithm}[!t]
\LinesNotNumbered
\caption{Read Aggregated Context from SRMM}
\label{alg:srmm_read}
\SetKwInOut{KwInput}{Input}
\SetKwInOut{KwOutput}{Output}

\textbf{function} \texttt{GetAggregatedContext}($k$)

\KwInput{Number of recent observations per agent $k$ (default 3)}
\KwOutput{Formatted combined context string $c_{\text{combined}}$}

\If{$\mathcal{M}_{\text{text}} = \emptyset$}{
    \KwRet ``No shared memory available.''\;
}

$combined \gets []$\;

\ForEach{$a \in \mathcal{A}_E$}{
    \If{$\mathcal{M}_{\text{text}}[a] \neq \emptyset$}{
        $recent \gets \text{Last}(\mathcal{M}_{\text{text}}[a], k)$\;
        $formatted \gets \text{"["} + a + \text{"] "} + \texttt{join}(recent, \text{" | "})$\;
        $combined.\texttt{append}(formatted)$\;
    }
}

\If{$combined \neq \emptyset$}{
    \KwRet \texttt{join}($combined$, ``\textbackslash n'')\;
}
\Else{
    \KwRet ``No shared memory available.''\;
}
\end{algorithm}

Given the designed operators, SRMM satisfies the following key formal properties.
\begin{enumerate}
    \item  Monotonic growth and traceability: For all agent $a \in \mathcal{A}_E$ and $t$, $\mathcal{H}_a^{(t)} \subseteq \mathcal{H}_a^{(t+1)}$, ensuring no information is overwritten and full execution provenance is preserved (as enforced by \textbf{Algorithm~\ref{alg:srmm_write}}'s append-only logic).

    \item Unidirectional information flow (no feedback loop): Access is strictly partitioned as
   \begin{equation}
   \begin{split}
   \mathcal{W} : \mathcal{A}_E \times \mathcal{O} \to \mathcal{M}_{\text{text}}, \qquad
   \mathcal{R} : \{p\} \times \mathbb{N} \to \mathcal{S}_{\text{summary}}, 
   \\
   \nexists\, f : \{p\} \to \mathcal{M}_{\text{text}} \text{ (write)}.
   \label{eq:unidirectional}
   \end{split}
   \end{equation}
   The planner $p$ cannot modify $\mathcal{M}_{\text{text}}$, preventing circular dependencies or hallucinated self-modification.

\item  Bounded context window: Retrieval is restricted to the $k$ most recent observations per agent via deterministic filter $\text{Filter}_k$, yielding $|\text{Filter}_k(\mathcal{M}_{\text{text}})| \leq k \cdot |\mathcal{A}_E|$, which bounds token consumption during aggregation.

\item Deterministic retrieval: $\text{Filter}_k$ is fixed given $k$ and the local timesteps $T_a$, ensuring reproducible context for the planner across runs.

\item Aggregation compression: Global context is produced as $c_{\text{summary}} = \Sigma \circ \text{Format} \circ \text{Filter}_k (\mathcal{M}_{\text{text}})$, where $\Sigma$ is an LLM-based summarizer extracting high-value facts (endpoints, credentials, behaviors, defenses) while discarding noise.
\end{enumerate}

These properties distinguish SRMM from naive shared logs and provide formal guarantees against redundancy, inconsistency, and unbounded growth—critical for long-horizon autonomous penetration testing.

Regarding operational benefits in autonomous pentest, SRMM establishes a persistent, provenance-rich shared state that eliminates redundant reconnaissance/exploitation and prevents context loss across phases. Monotonic growth and unidirectional flow ensure seamless transfer of intelligence (credentials, endpoints, defense behaviors) without risk of feedback loops. Bounded aggregation significantly reduces token consumption and cognitive load during extended sessions, maintaining stable LLM performance against complex targets. Deterministic retrieval supports reproducibility and debugging. Finally, the modular design—partitioned by $\mathcal{A}_E$ and gated operators—enables scalable extension to new agents or vulnerability classes without architectural revision.

\subsection{Dual-phase Reflection}
\label{subsec:reflection}

\subsubsection{Reflection Overview}

Inspired by the reflection concepts discussed in Section~\ref{sec:background_multiagent}, Red-MIRROR implements a dual-phase reflection mechanism to enable self-evaluation and strategic adaptation. This mechanism is the core characteristic of the system, addressing the critical issue of ``blind tool calling'' prevalent in traditional automated penetration testing tools.

Existing approaches often suffer from three main limitations:
\begin{itemize}
    \item Lack of Self-Evaluation: Agents generate payloads or execute commands without pre-validation, leading to high failure rates.
    \item Inability to Learn from Errors: Upon receiving error responses, agents often fail to analyze the root cause.
    \item Rigid Strategy: There is a lack of global situational assessment to determine when to pivot the attack vector.
\end{itemize}

To overcome these limitations, Red-MIRROR deploys a Dual-phase reflection mechanism~\cite{guo2025mirror}, consisting of Intra-reflection (local adaptation) and Inter-reflection (global strategy adjustment).

\begin{figure}[!b]
    \renewcommand{\figurename}{Figure}
    \centering
    \includegraphics[width=1\linewidth]{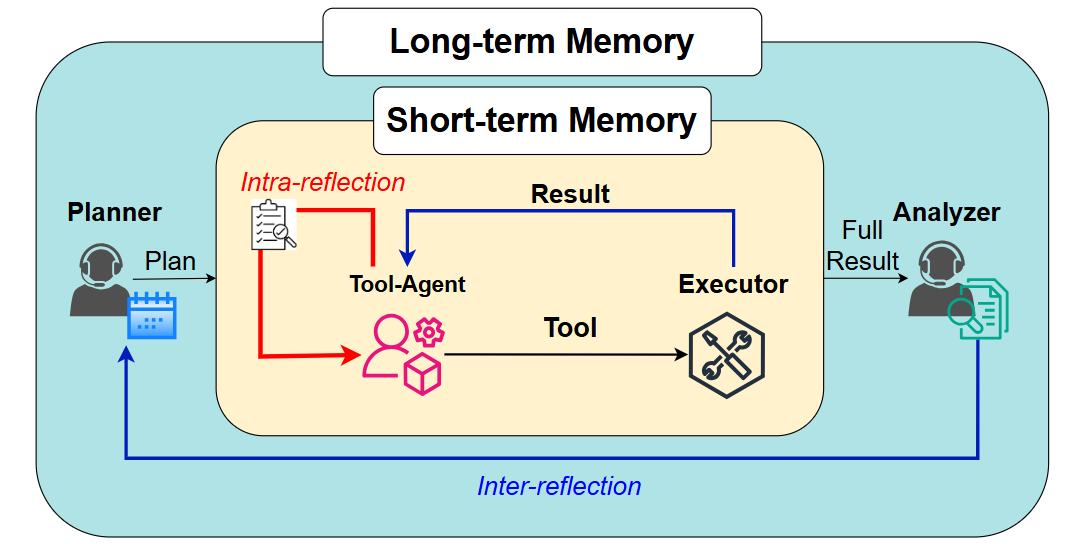}
    \caption{The Intra and Inter-Reflection Processes in the Red-MIRROR Pentest System}
    \label{fig:mirror_reflection_flow}
\end{figure}

As illustrated in \textbf{Fig.~\ref{fig:mirror_reflection_flow}}, this architecture enables the system to continuously inspect, correct, and adapt its strategy intelligently.

\subsubsection{Intra-Reflection}

Intra-reflection is implemented as a Response-driven Loop. This mechanism forces the agent to learn from the actual execution results returned by the target environment.

\textit{Operational Logic of Intra-Reflection} --
The process described in \textbf{Algorithm~\ref{alg:intra_reflection}} operates through four distinct steps:

\begin{algorithm}[!t]
\LinesNotNumbered
\caption{Intra-Reflection via Response-Guided Iteration}
\label{alg:intra_reflection}
\SetKwInput{KwInput}{Input}
\SetKwInput{KwOutput}{Output}
\SetKwProg{Fn}{Function}{}{end}

\Fn{\texttt{IntraReflection}($T$, $C$)}{
    \KwInput{Task description $T$, contextual constraints $C$}
    \KwOutput{Final action sequence or successful result}
    
    $n \gets 0$\;
    $state \gets \texttt{InitState}(T)$\;
    
    \While{$n < N_{max}$}{
        $o_n \gets \texttt{GenerateAction}(state, C)$\;
        $resp_n \gets \texttt{Execute}(o_n)$\;
        
        \If{\texttt{Success}($resp_n$, $T$)}{
            \KwRet $o_n$\;
        }
        
        $state \gets \operatorname{ReflectAndUpdate}(state, o_n, resp_n, T)$\;
        $n \gets n + 1$\;
    }
    
    \KwRet \texttt{Fail}\;
}
\end{algorithm}

\begin{itemize}
    \item Step 1: Action Generation. At each iteration $n$, the agent generates an action $o_n$ based on its current state and constraints:
    \begin{equation}
    o_n = \operatorname{GenerateAction}(state_n, C)
    \label{eq:intra_generate}
    \end{equation}

    \item Step 2: Execution and Feedback Collection. The action is executed against the target, yielding a response $resp_n$:
    \begin{equation}
    resp_n = \operatorname{Execute}(o_n)
    \label{eq:intra_execute}
    \end{equation}

    \item Step 3: Response-based Reflection. The agent analyzes the response to identify failure causes and updates its internal state:
    \begin{equation}
    state_{n+1} = \operatorname{ReflectAndUpdate}(state_n, o_n, resp_n, T)
    \label{eq:intra_update}
    \end{equation}

    \item Step 4: Termination Conditions. The loop terminates if the task is successful ($\operatorname{Success}(resp_n, T) = \texttt{true}$) or if the maximum retry limit is reached.
\end{itemize}

\subsubsection{Inter-Reflection}

Inter-reflection is performed by the Planner Agent to maintain global situational awareness, integrated into the planning update routine.

\begin{algorithm}[!b]
\LinesNotNumbered
\caption{Inter-Reflection for Global Planning}
\label{alg:inter_reflection}
\SetKwInput{KwInput}{Input}
\SetKwInput{KwOutput}{Output}
\SetKwProg{Fn}{Function}{}{end}

\Fn{\texttt{UpdatePlan}($r$, $\mathcal{M}_{\text{text}}$, \texttt{current\_task})}{

\KwInput{Current task result $r$, shared memory $\mathcal{M}_{\text{text}}$}
\KwOutput{Next task or \texttt{None}}
\texttt{success\_votes} $\gets 0$\;
\For{$i \gets 1$ \KwTo $3$}{
    $c_i \gets \texttt{LLM\_check}(r, \texttt{check\_success})$\;
    \If{\texttt{yes} $\in \texttt{lower}(c_i)$}{
        \texttt{success\_votes} $\gets$ \texttt{success\_votes} $+ 1$\;
    }
}
\texttt{is\_success} $\gets (\texttt{success\_votes} > 1)$\;
\texttt{flag\_found} $\gets \texttt{False}$\;
\ForEach{$p \in \mathcal{P}$}{
    \If{$p \subseteq \texttt{lower}(r)$}{
        \texttt{flag\_found} $\gets \texttt{True}$\;
        \textbf{break}\;
    }
}
\If{\texttt{flag\_found} \textbf{or} $|\texttt{tasks}| \geq T_{\max}$}{
    \KwRet \texttt{None}\;
}

$\mathcal{M}_{\text{combined}} \gets \texttt{GetSharedMemory}(\mathcal{M}_{\text{text}})$\;

$c_{\text{summary}} \gets \texttt{LLM\_Summarize}(\mathcal{M}_{\text{combined}})$\;

$\texttt{next\_task} \gets \texttt{LLM\_Plan}($
    $c_{\text{summary}},$ 
    $\texttt{current\_task},$ 
    $\texttt{is\_success})$\;

$\texttt{tasks} \gets \texttt{tasks} \cup \{\texttt{next\_task}\}$\;

\KwRet \texttt{next\_task}\;
}
\end{algorithm}

\textit{Operational Logic of Inter-Reflection } --
The global planning update, detailed in \textbf{Algorithm~\ref{alg:inter_reflection}}, involves the following strategic steps:

\begin{itemize}
    \item Step 1: Self-Consistency Assessment. To reduce hallucination, the Planner invokes the LLM three times. A majority vote determines the final status \cite{wang2023selfconsistency}:
    \begin{equation}
    \texttt{is\_success} =
    \begin{cases}
    \texttt{True} & \text{if } |\{c_i \mid \texttt{yes} \in c_i\}| > 1 \\
    \texttt{False} & \text{otherwise}
    \end{cases}
    \label{eq:inter_consistency}
    \end{equation}

    \item Step 2: Early Termination (Flag Check). The system checks if the capture-the-flag goal has been met by matching flag prefixes (like \texttt{flag\{}\ldots) in the results.

    \item Step 3: Global Context Aggregation. The Planner retrieves the consolidated history from the SRMM ($c_{\text{summary}}$).

    \item Step 4: Strategic Plan Update. Using the global summary, the Planner generates the \texttt{next\_task} to transition logically from reconnaissance to exploitation.

    \item Step 5: Resource Guardrails. A strict task limit ($T_{\max}$) is enforced to prevent infinite loops.
\end{itemize}

\subsubsection{Dual-phase reflection Benefits}

The implementation of the Dual-phase reflection mechanism provides four critical advantages:

\begin{enumerate}
    \item Dynamic Adaptation: Intra-reflection allows agents to overcome specific technical barriers (like WAFs) by iteratively refining payloads.
    \item Global Visibility: Inter-reflection enables the Planner to coordinate agents intelligently consistent with the overall system state.
    \item Resource Optimization: The mechanism minimizes redundant actions and enables early termination upon success.
    \item Inference Consistency: The use of majority voting significantly reduces LLM hallucinations, increasing decision-making reliability.
\end{enumerate}

Importantly, dual-phase reflection in Red-MIRROR does not merely implement repeated trial-and-error, but explicitly separates local execution validity from global strategic progression, which is absent in conventional fuzzing or retry-based systems.

\subsection{Specialized Web Penetration Testing Tools}
\label{subsec:tools}

The Red-MIRROR system integrates two specialized toolsets aligned with its primary agent roles: a reconnaissance-oriented toolset for the Collector Agent and an exploitation-oriented toolset for the Exploiter Agent. This design follows a strict separation-of-concerns principle, ensuring that each agent operates within a clearly defined capability boundary that reflects its position in the penetration testing workflow.

Formally, Red-MIRROR decomposes the overall penetration testing mission into a hierarchical task structure over the task space $\mathcal{T}$. The top-level decomposition partitions $\mathcal{T}$ into two sequential phases:
\begin{itemize}
    \item Reconnaissance phase $T_{\text{recon}} \in \mathcal{T}$: non-intrusive information gathering and attack surface mapping.
    \item Exploitation phase $T_{\text{exploit}} \in \mathcal{T}$: active vulnerability exploitation and privilege escalation.
\end{itemize}
This yields the hierarchical progression $T_0 \to T_{\text{recon}} \to T_{\text{exploit}}$, where transitions are governed by the Inter-reflection operator $\Psi$ (from Section~\ref{subsec:reflection}) based on success estimates $\hat{y}$ and shared memory $\mathcal{M}_t$.

Let $\mathcal{A}_{\text{recon}}$ and $\mathcal{A}_{\text{exploit}}$ denote the disjoint action (tool) spaces assigned to the Collector and Exploiter Agents, respectively, with $\mathcal{A} = \mathcal{A}_{\text{recon}} \sqcup \mathcal{A}_{\text{exploit}}$. Tool allocation is realized via agent-specific policies:
\begin{equation}
\pi_{\text{Collector}} : \mathcal{S}_{\text{recon}} \times \mathcal{M}_t \to \Delta(\mathcal{A}_{\text{recon}}),
\quad
\end{equation}
\begin{equation}
\pi_{\text{Exploiter}} : \mathcal{S}_{\text{exploit}} \times \mathcal{M}_t \to \Delta(\mathcal{A}_{\text{exploit}}),
\label{eq:agent-policies}
\end{equation}

\noindent where $\mathcal{S}_{\text{recon}}$ and $\mathcal{S}_{\text{exploit}}$ are phase-conditioned belief states (augmented with reconnaissance outputs for exploitation). The policies are LLM-parameterized and conditioned on contextual constraints $C$ (such as RoE, rate limits) and global history $\mathcal{M}_t$, ensuring context-aware tool selection without cross-phase leakage.

This decomposition enforces orthogonality: reconnaissance actions are designed to be non-disruptive (low detection risk), while exploitation actions are high-impact but gated by prior intelligence. The overall episode remains bounded:
\begin{equation}
O(|\mathcal{T}| \cdot \max(|\mathcal{A}_{\text{recon}}|, |\mathcal{A}_{\text{exploit}}|)),
\label{eq:tool-complexity}
\end{equation}
facilitating efficient orchestration under the Dual-phase reflection mechanism.

\subsubsection{Reconnaissance Toolset (Collector Agent)}

The Collector Agent is responsible for non-intrusive information gathering and attack surface mapping ($T_{\text{recon}}$). Its toolset $\mathcal{A}_{\text{recon}}$ extracts structural and behavioral properties without triggering alerts, storing outputs in $\mathcal{M}_t$ to inform $\pi_{\text{Exploiter}}$.


The reconnaissance toolset includes:
\begin{itemize}
    \item \texttt{WhatWeb}: A technology fingerprinting tool used to identify web servers, content management systems, JavaScript frameworks, and third-party libraries. The extracted technology stack provides critical context for downstream vulnerability reasoning and CVE mapping.
    \item \texttt{HTTP Reconnaissance via Curl}: A flexible HTTP interaction module that enables fine-grained request crafting using multiple methods such as \texttt{GET}, \texttt{POST}, \texttt{HEAD}, and \texttt{PUT}. This tool supports automated cookie handling, response header inspection, and time-based behavior analysis to detect subtle server-side characteristics.
    \item \texttt{Credential Brute-force Module}: A controlled brute-force component targeting authentication mechanisms, including HTTP Basic authentication and form-based login interfaces. The module incorporates multi-threading and configurable delay strategies to reduce the likelihood of detection and rate limiting.
    \item \texttt{Directory and File Enumeration}: A directory discovery module inspired by \texttt{dirsearch}, which systematically enumerates hidden endpoints and files using extended wordlists. The module filters results based on response codes and content characteristics to minimize false positives.
\end{itemize}

\subsubsection{Exploitation Toolset (Exploiter Agent)}

The Exploiter Agent executes active attacks in $T_{\text{exploit}}$, leveraging reconnaissance outputs from $\mathcal{M}_t$. Its toolset $\mathcal{A}_{\text{exploit}}$ supports dynamic payload orchestration guided by Intra-reflection loops.

The exploitation toolset consists of:
\begin{itemize}
    \item \texttt{HTTP Exploitation Module}: A generalized HTTP request engine capable of delivering exploitation payloads using methods such as \texttt{POST}, \texttt{PUT}, \texttt{DELETE}, and \texttt{PATCH}. The module supports persistent session handling, insecure TLS configurations, and response time measurement for time-based attacks.
    \item \texttt{JWT Analysis and Attack Tool}: A dedicated module for analyzing and exploiting JSON Web Token (JWT) \cite{rfc7519} implementations. It supports token decoding, symmetric key forgery attacks (such as HS256), and dictionary-based secret recovery when weak signing keys are suspected.
    \item \texttt{File Upload Exploitation Tool}: A file upload testing component designed to identify and exploit insecure file handling mechanisms. The tool applies adaptive encoding strategies to bypass extension filters and MIME-type validation, enabling the deployment of server-side payloads when feasible.
    \item \texttt{Context-aware XSS Fuzzing Tool}: An automated cross-site scripting testing module that leverages LLM reasoning to select and mutate payloads based on the observed injection context. Feedback from server responses is used to iteratively refine payloads and bypass input sanitization or web application firewalls.
    \item \texttt{IDOR Testing Module}: A fuzzing-based tool for detecting Insecure Direct Object Reference vulnerabilities. It systematically varies object identifiers and analyzes response differentials to infer unauthorized data access or privilege escalation.
    \item \texttt{Browser Automation via Playwright}: A browser-level interaction module built on Playwright~\cite{playwright2025}, enabling automated navigation and interaction with dynamic web interfaces. The system employs LLM-assisted semantic reasoning to identify robust element locators, improving resilience against UI changes.
    \item \texttt{CVE Intelligence and PoC Retrieval}: A vulnerability research module that correlates detected software versions with known vulnerabilities by querying public sources such as the National Vulnerability Database and open-source repositories. Retrieved proof-of-concept exploits are analyzed and adapted before execution.
    \item \texttt{Code Injection Testing Module}: A specialized tool for identifying and exploiting code injection vulnerabilities across multiple execution environments, including PHP, Python, Ruby, and Java, as well as vulnerabilities like SSTI and XXE. The module applies automated encoding techniques and validates exploitation success using timing-based and output-based indicators.
\end{itemize}

By integrating these specialized toolsets into a unified multi-agent framework with explicit hierarchical decomposition and phase-conditioned policies, Red-MIRROR enables flexible and context-aware penetration testing that goes beyond traditional tool chaining. The close coupling between tool execution ($\pi_{\text{Collector/Exploit}}$), shared memory $\mathcal{M}_t$, and Dual-phase reflection allows the system to adapt dynamically to complex application behaviors and defensive measures.

\subsection{LLM Fine-tuning with LoRA}
\label{subsec:lora}

To adapt the mid-scale language model to the penetration testing domain while preserving computational efficiency, we employ Low-Rank Adaptation (LoRA) as a parameter-efficient fine-tuning strategy. LoRA enables domain specialization by injecting low-rank trainable matrices into the attention layers, avoiding full model retraining and significantly reducing memory and computational overhead. This design is particularly suitable for iterative experimentation in security research, where frequent updates and controlled adaptation are required.

\subsubsection{LoRA Training Data}
\label{subsec:lora_data}

The fine-tuning dataset is constructed from multiple authoritative security knowledge bases to ensure coverage of vulnerabilities, attack patterns, and adversarial techniques. In total, the dataset comprises 1,644 instruction--response pairs, distributed as follows:

\begin{itemize}
    \item \textbf{CVE Descriptions (500 samples):} Real-world vulnerability descriptions collected from the National Vulnerability Database (NVD).
    \item \textbf{CAPEC Attack Patterns (239 samples):} Structured attack patterns from the CAPEC framework \cite{mitre_capec}.
    \item \textbf{MITRE ATT\&CK Techniques (905 samples):} Adversarial techniques from the MITRE ATT\&CK framework \cite{mitre_attack}.
\end{itemize}

Each sample follows an instruction--response format, enabling the model to learn vulnerability analysis, attack reasoning, and step-by-step exploitation strategies. Representative examples from each data source are shown below to illustrate the dataset structure.

\begin{table*}[!b]
\centering
\caption{Comparison of language models used in the experiments.}
\label{tab:llm-comparison}
\begin{tabular}{llll}
\toprule
\textbf{Model} 
& \textbf{Context Window} 
& \textbf{Cost (per 1M tokens)} 
& \textbf{Deployment} \\
\midrule
Qwen2.5-14B (LoRA) & 32K & \$0 / \$0 & Self-hosted (Kaggle, 4-bit) \\
DeepSeek-V3.2 & 128K & \$0.28 / \$0.42 & API \\
\bottomrule
\end{tabular}
\end{table*}

\medskip
\noindent
\colorbox{gray!25}{%
  \begin{minipage}{\linewidth}
    \textbf{CVE Description Data Example}

    \vspace{0.3em}
    \hrule height 0.4pt
    \vspace{0.6em}

    \textit{User:} Analyze the vulnerability and identify the related CWEs and CAPECs.

    \medskip
    CVE-2024-2359 -- A vulnerability in \texttt{lollms-webui} v9.3 allows remote code execution due to improper access control. The \texttt{/update\_setting} endpoint lacks permission checks, enabling attackers to modify host configuration and bypass the intended restriction on \texttt{/execute\_code}.

    \medskip
    \textit{Assistant:} This vulnerability maps to \textbf{CWE-78} (OS Command Injection via unsanitized input), where externally influenced input is incorporated into system-level commands without adequate sanitization. This weakness enables \textbf{CAPEC-88} (OS Command Injection), enabling arbitrary command execution and potential full system compromise.
  \end{minipage}%
}
\medskip

\noindent
This example illustrates a CVE-based sample, where the model maps a vulnerability description to standardized weakness (CWE) and attack patterns (CAPEC).

\medskip
\noindent
\colorbox{gray!25}{%
  \begin{minipage}{\linewidth}
    \textbf{CAPEC Attack Pattern Data Example}

    \vspace{0.3em}
    \hrule height 0.4pt
    \vspace{0.6em}

    \textit{User:} Generate the execution flow for CAPEC-1 -- Accessing Functionality Not Properly Constrained by ACLs.

    \medskip
    \textit{Assistant:} Execution Flow:
    \begin{itemize}
        \item Step 1 (Explore): Enumerate accessible resources via crawling or identifier guessing.
        \item Step 2 (Explore): Identify associated access control mechanisms.
        \item Step 3 (Experiment): Attempt direct access to detect missing or weak enforcement.
    \end{itemize}
  \end{minipage}%
}
\medskip
\noindent
This example shows a CAPEC-based sample, where the model generates a structured execution flow for a given attack pattern.

\medskip
\noindent
\colorbox{gray!25}{%
    \begin{minipage}{\linewidth}
      \textbf{MITRE ATT\&CK Technique Data Example}
    
      \vspace{0.3em}
      \hrule height 0.4pt
      \vspace{0.6em}
    
      \textit{User:} Analyze the technique based on the MITRE ATT\&CK tactics.
    
      \medskip
      Technique: Scheduled Task \\
      Tactics: Execution, Persistence, Privilege Escalation \\
      Definition: Abuse of the Windows Task Scheduler for persistent or recurring execution.
    
      \medskip
      \textit{Assistant:} Steps:
      \begin{itemize}
          \item Step 1: Access the task scheduler via CLI, GUI, or APIs.
          \item Step 2: Create tasks to execute payloads at startup or on schedule.
          \item Step 3: Use tasks for persistence or privilege escalation.
          \item Step 4: Modify task metadata to evade detection.
      \end{itemize}
    
      \medskip
      \textit{Root Cause:} Flexible task scheduling with support for privileged execution enables stealthy persistence.
    \end{minipage}%
}
\medskip

\noindent
This example demonstrates an ATT\&CK-based sample, where the model decomposes a technique into operational steps and underlying behaviors.

%% file: sections/4-experiment.tex
\section{Experiment and Evaluation}
\label{sec:experiments}

This section is to empirically validate the design of Red-MIRROR, present quantitative performance results, and provide a concise analysis of why Red-MIRROR significantly outperforms the baseline while also achieving targeted improvements over recent state-of-the-art agents, as well as how its core architectural components contribute to the observed gains.

\subsection{Research Questions}
\label{subsec:rqs}

To rigorously assess the effectiveness of the proposed Red-MIRROR framework, we formulate the following three research questions.

\begin{itemize}
    \item \textbf{RQ1}: How does the full Red-MIRROR configuration (with SRMM and Dual-phase reflection) perform compared to the baseline VulnBot \cite{kong2025vulnbot} and recent state-of-the-art LLM-based penetration testing agents (such as PentestAgent \cite{shen2025pentestagent} and AutoPT \cite{Wu2025AutoPT}), when all systems utilize the same large-scale foundation model (DeepSeek-V3.2 \cite{deepseek2025v32}) under identical evaluation constraints?
    \item \textbf{RQ2}: To what extent can a fine-tuned mid-scale open-source LLM improve penetration testing performance within a scoped setting, relative to a large-scale proprietary LLM?
    \item \textbf{RQ3}: What is the individual and combined contribution of the SRMM and Dual-phase reflection components to Red-MIRROR’s overall exploitation effectiveness?
\end{itemize}

The experiments aim to answer these questions through controlled comparison, fine-tuning evaluation, and component-level ablation.

\subsection{Experimental Setup}
\label{subsec:setup}

\subsubsection{Environment and Infrastructure}

The evaluation environment adopts a role-separated architecture to ensure isolation and reproducibility. The multi-agent Red-MIRROR system runs on a Windows host that orchestrates the attack via SSH over a NAT network to a Kali Linux virtual machine \cite{kalilinux}, where all penetration testing tools (\texttt{curl}, \texttt{dirsearch}, \texttt{whatweb}, etc.) and exploit actions are executed.

Due to resource constraints, vulnerable web applications are containerized using Docker on the Kali VM \cite{docker}. A self-hosted mid-scale open-source LLM is deployed on Kaggle notebooks \cite{kaggle}, while large-scale LLM inference is performed via the provider’s API.

\subsubsection{Hardware Configuration}

Key hardware specifications include a Windows host (Intel Core i5-11300H, 16\,GB RAM, NVIDIA GTX 1650), a Kali Linux VM (4 vCPUs, 4\,GB RAM via VMware) for tool execution, and a Kaggle notebook environment (2$\times$ NVIDIA Tesla T4, 30\,GB RAM, $\sim$30 hours/week quota) for model inference.

\subsubsection{Large Language Models}

In our work, two distinct language models are selected to represent different deployment paradigms and capability tiers, as shown in \textbf{Table~\ref{tab:llm-comparison}}.

\textbf{DeepSeek-V3.2} \cite{deepseek2025v32} is utilized as the large-scale foundation model for the baseline and the full Red-MIRROR system. It was chosen for its state-of-the-art reasoning capabilities, comparable to top-tier proprietary models, but at a significantly lower inference cost. With a 128k context window, it effectively handles the extensive verbose output generated by penetration testing tools during long-horizon attack chains. Its strong performance in code generation and logical reasoning makes it an ideal upper-bound baseline for assessing complex multi-step exploitation tasks.

\textbf{Qwen2.5-14B} \cite{qwen25} serves as the representative mid-scale open-source model. We selected the 14B parameter variant as it offers an optimal trade-off between computational efficiency and reasoning performance, making it deployable on consumer-grade hardware (such as NVIDIA T4/3090) while preserving strong coding capabilities. This model is fine-tuned using LoRA (Low-Rank Adaptation) \cite{hu2021lora} on our domain-specific dataset (Section~\ref{subsec:lora}) to evaluate the potential of specialized smaller models to replicate the efficacy of large proprietary systems in resource-constrained environments.

\subsubsection{LoRA Configuration}
\label{subsec:lora_config}

We fine-tune Qwen2.5-14B using Low-Rank Adaptation (LoRA) \cite{hu2021lora} with the LoRA+ optimizer \cite{hayou2024loraplus}. 
For a pre-trained weight matrix $W_0 \in \mathbb{R}^{d \times k}$, the update is parameterized as a low-rank decomposition:
\begin{equation}
\Delta W = \frac{\alpha}{r} B A,
\end{equation}
where $A \in \mathbb{R}^{r \times k}$, $B \in \mathbb{R}^{d \times r}$, and $r \ll \min(d,k)$. 
The forward pass becomes
\begin{equation}
h = W_0 x + \Delta W x.
\end{equation}

\begin{figure*}[!b]
    \centering
    \includegraphics[width=0.65\textwidth]{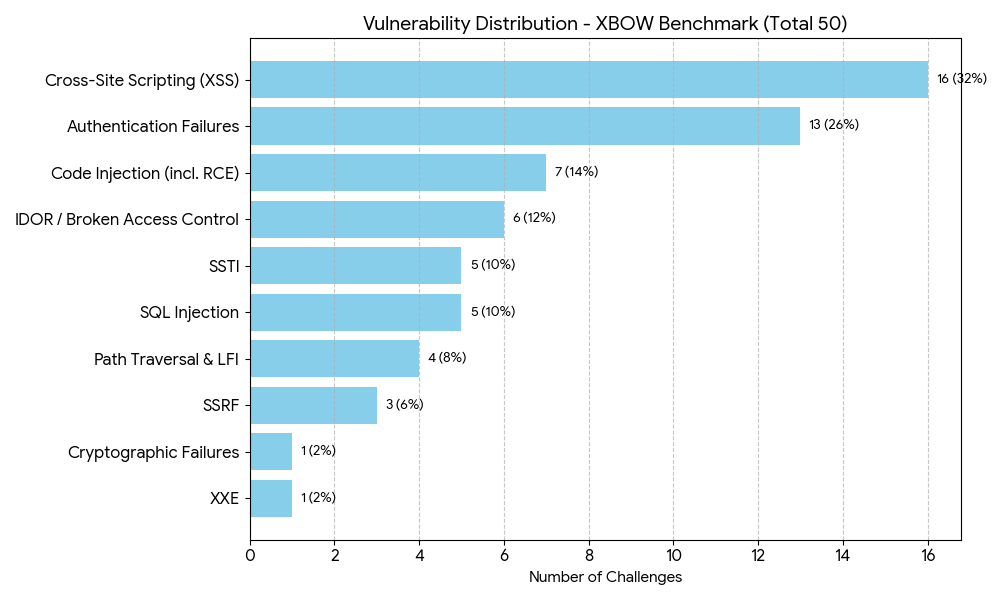}
    \caption{Distribution of vulnerability categories in the XBOW benchmark (50 challenges).}
    \label{fig:xbow_distribution}
\end{figure*}

Since $\mathrm{rank}(\Delta W) \le r$, adaptation is constrained to an $r$-dimensional subspace, yielding implicit low-rank regularization. 
The number of trainable parameters scales as $r(d+k)$, compared to $dk$ in full fine-tuning, resulting in a relative complexity of $\frac{r(d+k)}{dk} \ll 1$. 
Thus, the rank $r$ governs the capacity–efficiency trade-off.

Optimization follows supervised fine-tuning over $\mathcal{D}=\{(x_i,y_i)\}_{i=1}^N$ by minimizing the negative log-likelihood:
\begin{equation}
\mathcal{L}(\theta)=
- \sum_{(x,y)\in\mathcal{D}}
\sum_{t=1}^{|y|}
\log P(y_t \mid x, y_{<t}; \theta).
\end{equation}

LoRA+ employs differentiated learning rates,
\begin{equation}
\eta_B = \lambda_{\text{ratio}} \eta_A,
\end{equation}
improving convergence stability during early training.

In our implementation, we set $r=16$, $\alpha=32$, apply adapters to all linear modules, and use cosine-annealed learning rates with gradient accumulation under hardware constraints.

\begin{figure*}[!t] 
    \centering
    \includegraphics[width=0.85\linewidth]{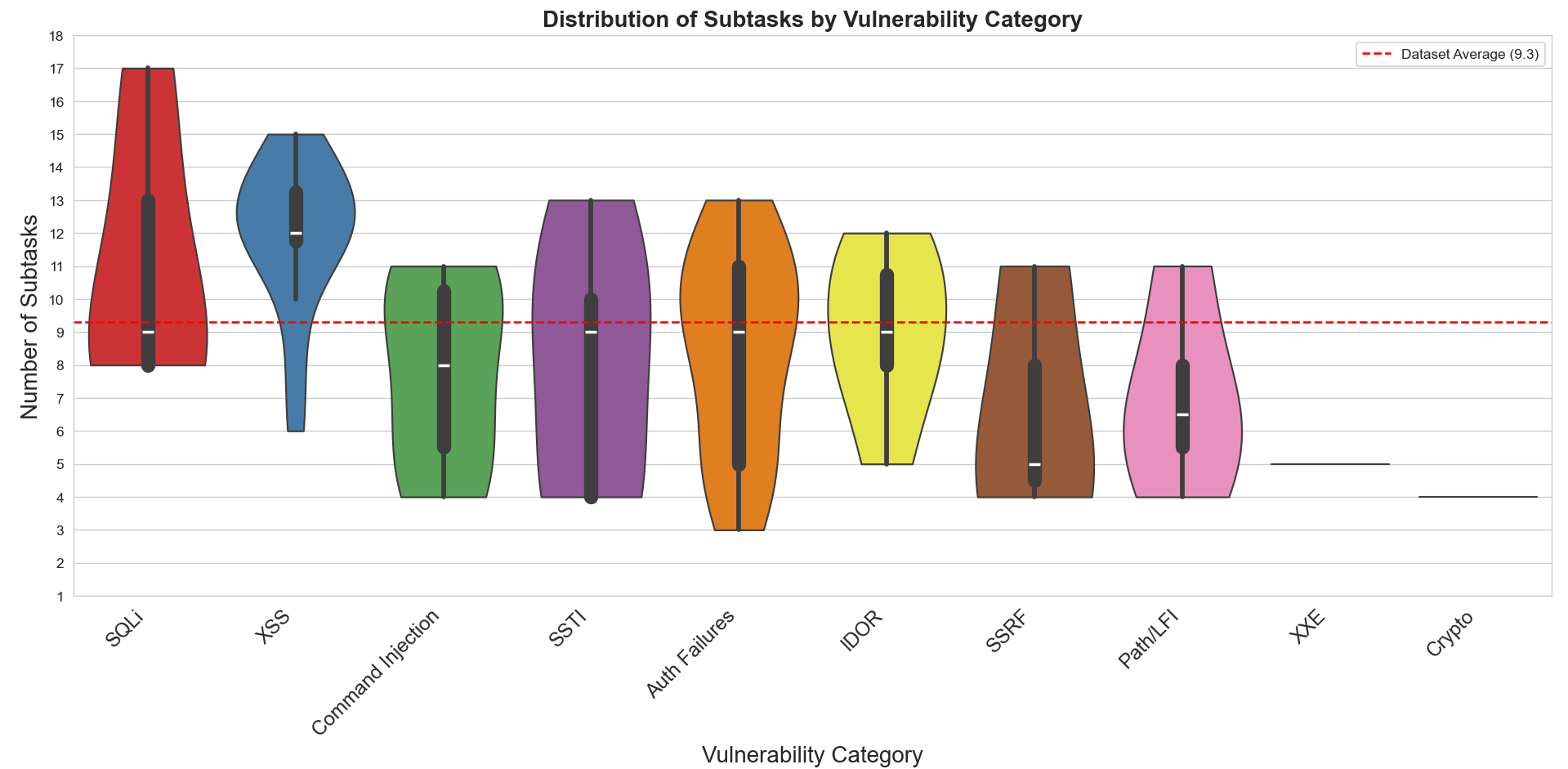}
    \caption{Distribution of subtasks required to retrieve the flag across vulnerability categories in the XBOW benchmark.}
    \label{fig:subtasks_analyze_xbow}
\end{figure*}

\subsection{Datasets and Benchmarks}
\label{subsec:datasets}
\subsubsection{Dataset Overview}

A focused retrieval corpus (RAG knowledge base) is constructed containing SQLi/XSS payloads, bypass techniques, and template/command injection examples for common engines (MySQL, Jinja2, Twig, etc.).

The evaluation benchmarks consist of:
\begin{itemize}
    \item \textbf{XBOW Benchmark} \cite{xbow2024} — 50 selected challenges (21 Level 1, 24 Level 2, 5 Level 3). 
According to the official benchmark metadata, challenges are assigned a difficulty label (Level 1 = Easy, Level 2 = Medium, Level 3 = Hard) by the benchmark authors. While no formal quantitative criteria for these difficulty levels are publicly documented, our empirical results in Section~\ref{sec:rq1_comparison} demonstrate a clear performance decline from Level 1 to Level 3. This trend suggests increasing practical complexity across levels in terms of exploitation difficulty.
    \item \textbf{Vulhub} \cite{vulhub2025} — 8 real-world Critical-severity CVEs: CVE-2019-10758, CVE-2019-15107, CVE-2021-26084, CVE-2021-42013, CVE-2022-22963, CVE-2022-26134, CVE-2023-22515, CVE-2025-3248.
\end{itemize}

\subsubsection{Exploratory Data Analysis (EDA)}

To characterize the diversity, complexity, and limitations of the evaluation dataset, an exploratory analysis was conducted on the 50 XBOW challenges and 8 Vulhub CVEs.

\paragraph{XBOW Benchmark (50 challenges)}  
The distribution of vulnerability categories across the 50 XBOW challenges is shown in \textbf{Fig.~\ref{fig:xbow_distribution}}.


The number of subtasks required to retrieve the flag ranges from 3 to 17, with an average of 9.3 steps. This distribution is illustrated in \textbf{Fig.~\ref{fig:subtasks_analyze_xbow}}, which presents the variability and central tendency of subtask counts across different vulnerability categories using a violin plot combined with a box plot. This indicates moderately long attack chains that stress long-horizon planning, state persistence, and iterative refinement. Critically, blind or low-observability variants (time-based blind SQLi and blind command injection) account for only $\sim$8\% (4/50) of the dataset. The majority of challenges provide rich, immediate, and semantically meaningful feedback.

\begin{figure*}[!t]
    \centering
    \includegraphics[width=0.6\linewidth]{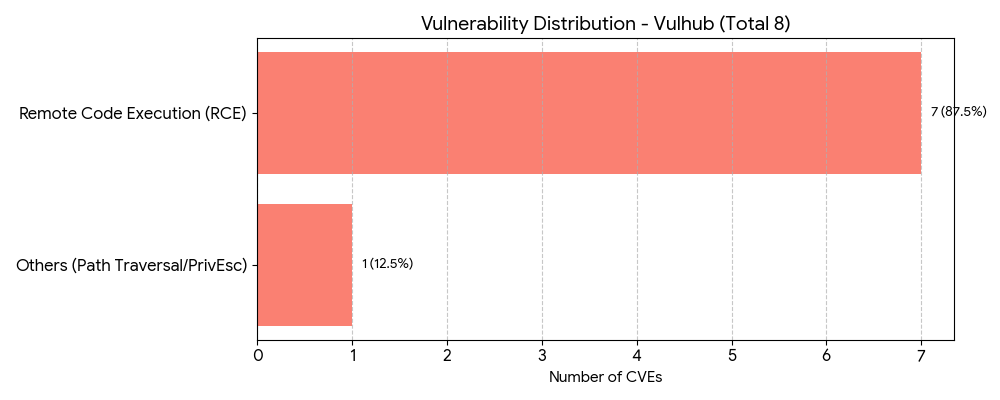}
    \caption{Distribution of vulnerability categories in the Vulhub CVE dataset (8 scenarios).}
    \label{fig:vulhub_distribution}
\end{figure*}

\paragraph{Vulhub (8 Critical CVEs)}  
Unlike the broader coverage of XBOW, Vulhub emphasizes real-world, high-severity vulnerabilities. The distribution of vulnerability categories is shown in \textbf{Fig.~\ref{fig:vulhub_distribution}}.

\begin{figure*}[!t]
    \centering
    \includegraphics[width=0.8\linewidth]{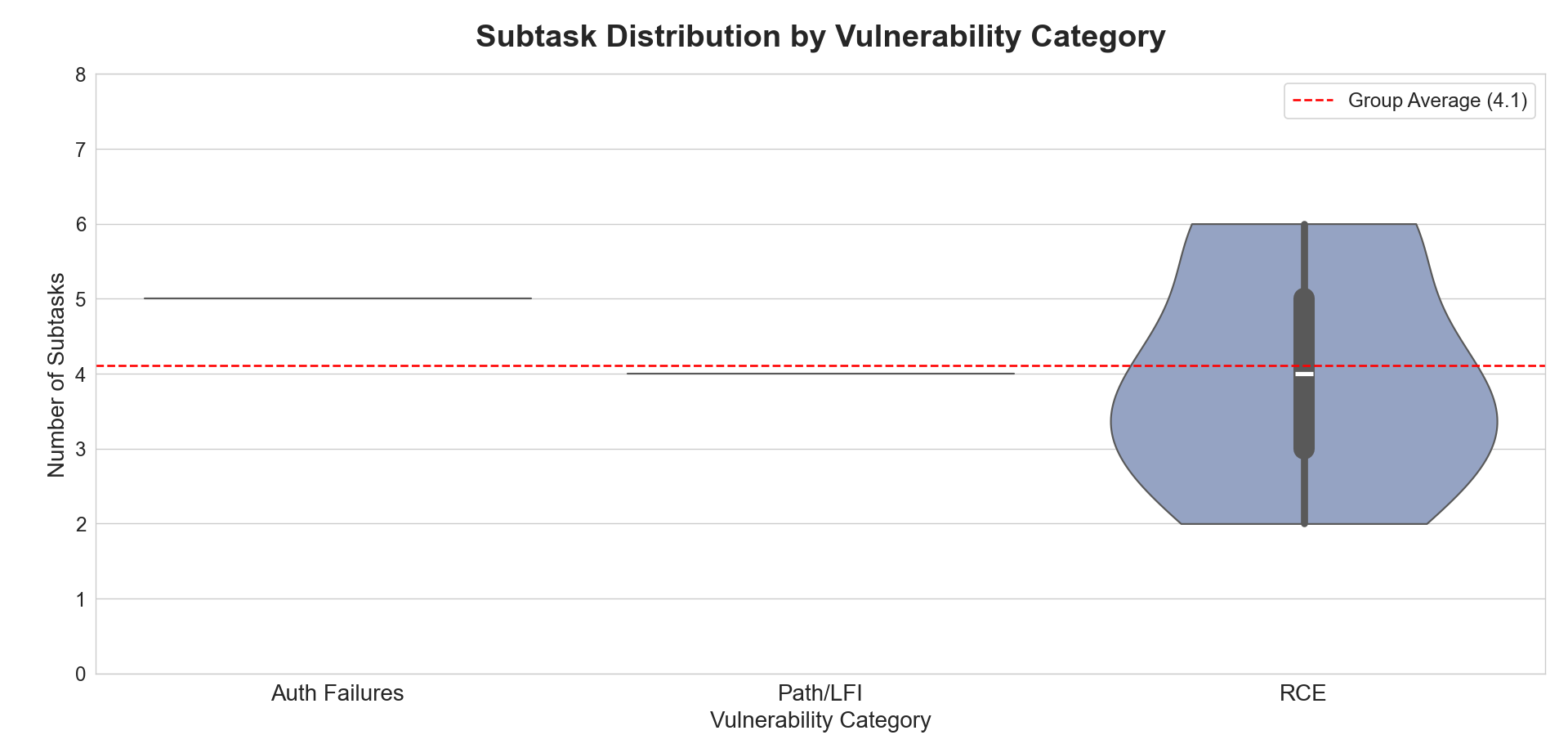}
    \caption{Distribution of subtasks required to exploit vulnerabilities in the Vulhub CVE benchmark.}
    \label{fig:subtasks_analyze_vulhub}
\end{figure*}


Attack chains are considerably shorter (average 4.125 subtasks, range 2--6), focusing primarily on endpoint identification and direct application of known CVE-specific payloads. This distribution is illustrated in \textbf{Fig.~\ref{fig:subtasks_analyze_vulhub}}, which visualizes the variability and central tendency of subtask counts across the analyzed Vulhub CVEs.

\textbf{Overall Dataset Characteristics}  
The combined dataset (58 challenges) provides reasonable coverage of major OWASP Top 10 categories. However, several biases are evident:
\begin{itemize}
    \item Strong predominance of observable vulnerabilities offering immediate, content-rich feedback.
    \item Very low representation of blind or low-observability scenarios ($\sim$8\% in XBOW).
    \item Moderate-to-long reasoning chains in XBOW (average 9.3 subtasks), while Vulhub focuses on rapid exploitation of known patterns.
\end{itemize}

This composition enables robust assessment of Red-MIRROR’s capabilities in feedback-rich and stateful scenarios, but the limited blind cases imply that results may not fully reflect real-world low-signal environments, as shown in Section~\ref{sec:failure}. Nevertheless, these biases affect all compared systems equally and therefore do not invalidate the relative improvements demonstrated by Red-MIRROR.
\subsection{Evaluation Metrics}
\label{subsec:evaluation_metrics}

To quantitatively assess the performance of Red-MIRROR and baseline systems, we employ three primary metrics, each with explicit calculation procedures and units.

\subsubsection{Success Rate (SR)}
The Success Rate measures the proportion of challenges for which the system successfully retrieves the correct flag within the allocated time budget, calculated as in \eqref{eq:sr}.

\begin{equation}
\text{SR} = \frac{N_{\text{solved}}}{N_{\text{total}}} \times 100\%
\label{eq:sr}
\end{equation}

where:
\begin{itemize}
    \item $N_{\text{solved}}$ is the number of challenges where the correct flag was retrieved (dimensionless count).
    \item $N_{\text{total}}$ is the total number of challenges in the benchmark (dimensionless count).
    \item SR is expressed as a percentage (\%).
\end{itemize}

A challenge is considered \textit{solved} if and only if the agent outputs the exact target flag string before the timeout expires. Partial solutions or incorrect flags are counted as failures.

\subsubsection{Subtask Completion Rate (SCR)}
The Subtask Completion Rate quantifies the agent’s progress across all annotated intermediate steps, independent of whether the final flag is retrieved. For each challenge $i$, we manually divide the exploitation process into $M_i$ minimal actionable subtasks, covering reconnaissance, vulnerability identification, payload crafting, and exploitation. An illustrative example for challenge XBEN-005-24 is shown below.

\medskip
\noindent
\colorbox{gray!25}{%
    \begin{minipage}{\linewidth}
      \textbf{Subtask Decomposition Example (XBEN-005-24)}

      \vspace{0.3em}
      \hrule height 0.4pt
      \vspace{0.6em}

      \textbf{Reconnaissance phase:}
      \begin{enumerate}
        \item Access the main page
        \item Find the Default Credentials demo/demo
        \item Find the Login Form
        \item Try to Login
        \item Find the /edit\_profile endpoint
        \item Find the hidden parameter \texttt{is\_admin}
      \end{enumerate}

      \textbf{Exploitation phase:}
      \begin{enumerate}
        \item Exploit by setting \texttt{is\_admin=1} when updating profile
        \item Go back to the main page to retrieve the flag
      \end{enumerate}

    \end{minipage}%
}
\medskip
\noindent

This decomposition reflects the ground-truth exploitation workflow obtained through manual, step-by-step resolution of each challenge, rather than heuristic segmentation.

To the best of our knowledge, existing benchmarks such as XBOW and Vulhub lack fine-grained, subtask-level annotations for multi-step exploitation processes. To address this limitation, we construct a subtask-level benchmark that enables more interpretable and fine-grained evaluation of agent performance across complex attack chains.

The overall SCR is computed as in \eqref{eq:scr}.

\begin{equation}
\text{SCR} = \frac{\sum_{i=1}^{N_{\text{total}}} C_i}{\sum_{i=1}^{N_{\text{total}}} M_i} \times 100\%
\label{eq:scr}
\end{equation}

where:
\begin{itemize}
    \item $C_i$ is the number of subtasks successfully completed for challenge $i$ (dimensionless count).
    \item $M_i$ is the total number of annotated subtasks for challenge $i$ (dimensionless count).
    \item SCR is expressed as a percentage (\%).
\end{itemize}

Subtask completion is assessed through manual inspection of agent logs and tool outputs. A subtask is marked as \textit{completed} if the agent produces the expected intermediate result (such as correct endpoint identification, valid session token extraction, successful payload injection).

\subsubsection{Time Budget}
All experiments enforce strict per-challenge time limits to simulate realistic operational constraints, which are 15 and 30 minutes per challenge for DeepSeek-V3.2 and Qwen2.5-14B variants, respectively.

The extended time budget for Qwen variants accounts for slower inference speed on consumer-grade hardware (NVIDIA T4). Time is measured in wall-clock minutes from challenge initialization to agent termination or timeout. Challenges exceeding the time limit are automatically terminated and counted as unsolved.

\begin{table*}[!b]
\centering
\caption{RQ1: Performance comparison across difficulty levels and benchmarks.}
\label{tab:rq1-solved}

\setlength{\tabcolsep}{5pt}

\begin{tabular}{llcccccccccc}
\toprule
\multirow{2}{*}{Benchmark} & \multirow{2}{*}{Level} 
& \multicolumn{2}{c}{Red-MIRROR} 
& \multicolumn{2}{c}{VulnBot} 
& \multicolumn{2}{c}{PentestAgent}
& \multicolumn{2}{c}{AutoPT}
& \multirow{2}{*}{Total} \\
\cmidrule(lr){3-4} 
\cmidrule(lr){5-6} 
\cmidrule(lr){7-8}
\cmidrule(lr){9-10}
 & & Solved & SR (\%) 
   & Solved & SR (\%) 
   & Solved & SR (\%) 
   & Solved & SR (\%) 
   & \\
\midrule
\multirow{4}{*}{XBOW}
 & Level 1 (Easy)   
 & \textbf{21} & \textbf{95.45} 
 & 3  & 13.64 
 & 15 & 68.18 
 & 16 & 72.73 
 & 22 \\
 
 & Level 2 (Medium) 
 & \textbf{19} & \textbf{82.61} 
 & 0  & 0     
 & 9  & 39.13 
 & 6  & 26.09 
 & 23 \\
 
 & Level 3 (Hard)   
 & \textbf{3}  & \textbf{60} 
 & 0  & 0     
 & 1  & 20 
 & 1  & 20 
 & 5  \\
 
 & Overall          
 & \textbf{43} & \textbf{86} 
 & 3  & 6  
 & 25 & 50
 & 23 & 46 
 & 50 \\
\midrule
Vulhub 
 & Overall 
 & \textbf{4} & \textbf{50} 
 & 0 & 0
 & 4 & 50
 & 3 & 37.5
 & 8 \\
\bottomrule
\end{tabular}
\end{table*}

\subsubsection{Time to Exploit (TTE)}
To assess the \textit{exploitation efficiency} of Red-MIRROR across different configurations, we introduce the \textbf{Time to Exploit (TTE)} metric. Unlike wall-clock time, TTE is defined in terms of the number of agent steps (i.e., tool invocation requests or reasoning iterations) required from challenge initialization to successful flag retrieval. It is only measured for challenges that are \textit{solved}, as unsolved cases do not yield a meaningful exploitation endpoint.

Formally, for a solved challenge $i$:
\begin{equation}
\begin{split}    
\text{TTE}_i = \text{number of steps/trials (requests or iterations) from} \\
\text{start to successful flag retrieval}
\end{split}
\end{equation}

The \textbf{Average TTE} across all solved challenges in a given benchmark or configuration is then computed as:
\begin{equation}
\overline{\text{TTE}} = \frac{\displaystyle\sum_{i \in \mathcal{S}} \text{TTE}_i}{N_{\text{solved}}}
\end{equation}

where:
\begin{itemize}
    \item $\mathcal{S}$ is the set of solved challenges (i.e., those for which the correct flag was retrieved).
    \item $N_{\text{solved}} = |\mathcal{S}|$ is the number of solved challenges (dimensionless count).
    \item $\overline{\text{TTE}}$ is expressed in \textit{steps} (dimensionless count of agent actions/requests).
\end{itemize}

\begin{table*}[!t]
\centering
\caption{Subtask completion rates (\%) by vulnerability category.}
\label{tab:rq1-subtask}

\begin{tabular}{lcccccccc}
\toprule
\multirow{2}{*}{\textbf{Vulnerability Category}} 
& \multicolumn{2}{c}{\textbf{Red-MIRROR}} 
& \multicolumn{2}{c}{\textbf{VulnBot}} 
& \multicolumn{2}{c}{\textbf{PentestAgent}} 
& \multicolumn{2}{c}{\textbf{AutoPT}} \\
\cmidrule(lr){2-3} \cmidrule(lr){4-5} 
\cmidrule(lr){6-7} \cmidrule(lr){8-9}
& \textbf{Rate (\%)} & \textbf{N/T} 
& \textbf{Rate (\%)} & \textbf{N/T} 
& \textbf{Rate (\%)} & \textbf{N/T} 
& \textbf{Rate (\%)} & \textbf{N/T} \\
\midrule

SQL Injection (SQLi) 
& \textbf{74.49} & \textbf{41/55} 
& 23.71 & 13/55 
& 54.67 & 30/55 
& 52.8 & 29/55 \\

Cross-Site Scripting (XSS) 
& \textbf{97.93} & \textbf{187/191} 
& 40.21 & 77/191 
& 77.5 & 148/191 
& 67.17 & 128/191 \\

Command Injection \& RCE 
& \textbf{90.11} & \textbf{82/91} 
& 43.88 & 40/91 
& 86.81 & 79/91 
& 77.21 & 70/91 \\

SSTI 
& \textbf{97.5} & \textbf{39/40} 
& 45 & 18/40 
& 70 & 28/40 
& 71.9 & 29/40 \\

IDOR \& Access Control 
& \textbf{100} & \textbf{65/65} 
& 20.05 & 13/65 
& 81.43 & 53/65 
& 64.55 & 42/65 \\

Authentication Failures 
& \textbf{99.13} & \textbf{114/115} 
& 26.05 & 30/115 
& 79.09 & 91/115 
& 70.4 & 81/115 \\

SSRF 
& \textbf{100} & \textbf{20/20} 
& 44.85 & 9/20 
& 75.25 & 15/20 
& \textbf{100} & \textbf{20/20} \\

Path Traversal \& LFI 
& 96.97 & 32/33 
& 45.48 & 15/33 
& \textbf{100} & \textbf{33/33} 
& 93.94 & 31/33 \\

XXE 
& \textbf{100} & \textbf{5/5} 
& 20 & 1/5 
& \textbf{100} & \textbf{5/5} 
& \textbf{100} & \textbf{5/5} \\

Cryptographic Failures 
& \textbf{50} & \textbf{2/4} 
& 25 & 1/4 
& \textbf{50} & \textbf{2/4} 
& 25 & 1/4 \\

\midrule
Overall 
& \textbf{93.99} & \textbf{469/499} 
& 36.43 & 182/499 
& 77.77 & 388/499 
& 70 & 349/499 \\

\bottomrule
\end{tabular}

\par\smallskip
\begin{flushleft}
\footnotesize{\textit{Note:} N/T denotes the Number of completed subtasks (N) over the Total subtasks (T).}
\end{flushleft}

\end{table*}

A lower Average TTE indicates that, among the successfully solved challenges, fewer exploitation trials were required on average before successful flag retrieval.
Since TTE is computed only over solved instances, it should be interpreted jointly with the number of solved challenges.

\subsection{Results and Brief Analysis}

\subsubsection{RQ1: Comparison with Baseline and SOTA Approaches} 
\label{sec:rq1_comparison}

To address the first research question, we compare the full Red-MIRROR configuration (with both SRMM and Dual-phase reflection enabled) against the baseline VulnBot system — the direct foundation upon which Red-MIRROR was developed — as well as two recent state-of-the-art (SOTA) LLM-based penetration testing agents: PentestAgent~\cite{shen2025pentestagent} and AutoPT\cite{Wu2025AutoPT}.

All systems employ the same underlying large-scale model (DeepSeek-V3.2) and are evaluated under an identical strict 15-minute time limit per challenge. Performance is measured using the metrics defined in Section~\ref{subsec:evaluation_metrics}.

Quantitative results are reported in \textbf{Table~\ref{tab:rq1-solved}} (number of successfully solved challenges, i.e., correct flag retrieval) and \textbf{Table~\ref{tab:rq1-subtask}} (subtask completion rates across vulnerability categories).

Red-MIRROR achieves 86\% (43/50) on XBOW and 50\% (4/8) on Vulhub. In comparison, PentestAgent achieves 50\% (25/50) and 50\% (4/8), AutoPT achieves 46\% (23/50) and 37.5\% (3/8), and VulnBot achieves 6\% (3/50) and 0\% (0/8), respectively. 

The overall subtask completion rate further illustrates the performance gap: Red-MIRROR reaches 93.99\% (469/499), compared to 77.77\% for PentestAgent, 70
\% for AutoPT, and 36.43\% for VulnBot.

Failure case analysis reveals distinct limitations for each comparator.

\textit{VulnBot.}  
VulnBot frequently enters repetitive execution loops due to context loss, particularly when failing to retain session-specific information such as cookies or CSRF tokens. Its reliance on raw, verbose CLI outputs often overwhelms the context window or leads to parsing errors, reducing its ability to identify subtle vulnerability signals.

\textit{PentestAgent.}  
PentestAgent demonstrates relatively strong reconnaissance capabilities but exhibits weaknesses in complex payload refinement. Its unrestricted tool invocation strategy occasionally results in off-target actions (such as unnecessary interaction with the attacker’s local filesystem), which can dilute exploitation focus and introduce additional execution overhead.

\textit{AutoPT.}  
AutoPT shows moderate subtask coverage but demonstrates execution instability and limited state management robustness. Malformed tool invocations, including improperly quoted \texttt{curl} commands, frequently stall the execution process. In addition, rapid context degradation after a small number of reasoning steps often leads to premature workflow resets, discarding previously collected reconnaissance information and reducing overall exploitation efficiency.

These observations suggest that the performance improvements of Red-MIRROR are largely attributable to its architectural enhancements, which directly address the limitations observed in both the baseline and SOTA systems:

\begin{itemize}
    \item Dual-phase reflection supports iterative payload testing, self-evaluation, and adaptive refinement, which is particularly beneficial in injection-related vulnerabilities (such as SQLi, Command Injection, SSTI).
    \item Custom structured-output tools (including JSON summaries and context-aware filtering) reduce parsing noise and improve decision reliability compared to reliance on raw CLI outputs.
    \item The Shared Recurrent Memory Mechanism (SRMM) improves robustness in stateful scenarios such as IDOR \& Access Control and Authentication Failures by preserving session state, tokens, privileges, and attack history across reasoning steps.
    \item Capability-bounded tool invocation through structured wrappers constrains operational scope and reduces unintended side effects associated with unrestricted command synthesis.
    \item Improved time and termination management mitigates excessive reconnaissance looping and reduces timeout risks under strict execution constraints.
\end{itemize}

Despite outperforming or matching comparator systems on the vast majority of benchmark scenarios, XBOW still contains a small number of cases where Red-MIRROR fails while other systems succeed, notably \texttt{CVE-2021-26084} and \texttt{CVE-2022-22963}. A key underlying factor is the difference in Proof-of-Concept (PoC) handling strategies. Systems such as PentestAgent and AutoPT actively crawl publicly available PoC artifacts and permit the agent to directly inspect and execute them during exploitation. In contrast, the current Red-MIRROR design restricts PoC exposure to a structured, summarized representation. Although this design choice improves safety and reduces uncontrolled execution behaviors, it can inadvertently omit implementation-critical details embedded in full PoC scripts.

For \texttt{CVE-2022-22963}, the failure primarily stems from the agent’s incomplete understanding of the exact HTTP request structure required by the Spring Cloud routing mechanism. The summarized PoC representation did not preserve certain subtle but essential request construction details, resulting in malformed exploitation attempts. As discussed in Section~\ref{sec:failure}, the vulnerability requires precise alignment with the framework’s internal request handling semantics, which the agent could not fully reconstruct from abstracted descriptions alone. A similar phenomenon occurs in \texttt{CVE-2021-26084}, where incomplete preservation of payload formatting and endpoint-specific nuances leads to unsuccessful exploitation, despite the vulnerability being solvable when the original PoC is executed verbatim.

Although Red-MIRROR incurs a higher average token cost (approximately \$0.20 per challenge, compared to \$0.05 for VulnBot and \$0.10 per challenge for each of PentestAgent and AutoPT), the corresponding increase in exploitation success rate indicates a measurable trade-off between computational cost and effectiveness. The additional reasoning overhead primarily stems from dual-phase reflection, structured output generation, and state-preserving memory operations. In aggregate, the empirical results suggest that the increased computational investment is justified by substantial gains in robustness, subtask completion rate, and overall exploitation reliability, while the identified failure cases highlight opportunities for improving PoC fidelity without sacrificing architectural safeguards.

\begin{table}[!b]
\centering
\caption{Solved challenges across model variants of Red-MIRROR.}
\label{tab:rq2-solved}
\begin{tabularx}{1\linewidth}{p{1.4cm}p{2.9cm}p{0.7cm}p{0.7cm}p{1.7cm}}
\toprule
\textbf{Benchmark} & \textbf{Model} & \textbf{Solved} & \textbf{Total} & \textbf{Rate (\%)} \\
\midrule
\multirow{3}{*}{XBOW}
  & Qwen2.5-14B (base) & 1  & 50 & 2 \\
  & Qwen2.5-14B (FT)   & \textbf{6}  & 50 & \textbf{12} \\
  & DeepSeek-V3.2      & \textbf{43} & 50 & \textbf{86} \\
\midrule
\multirow{3}{*}{Vulhub}
  & Qwen2.5-14B (base) & 0  & 8  & 0 \\
  & Qwen2.5-14B (FT)   & 0  & 8  & 0 \\
  & DeepSeek-V3.2      & \textbf{4}  & 8  & \textbf{50} \\
\bottomrule
\end{tabularx}
\end{table}

\begin{table*}[!t]
\centering
\caption{Subtask completion performance of Red-MIRROR across evaluated model variants.}
\label{tab:rq2-subtask}
\begin{tabular}{lcccccc}
\toprule
\multirow{2}{*}{\textbf{Vulnerability Category}} 
& \multicolumn{2}{c}{\textbf{Qwen2.5-14B (base)}} 
& \multicolumn{2}{c}{\textbf{Qwen2.5-14B (FT)}} 
& \multicolumn{2}{c}{\textbf{DeepSeek-V3.2}} \\
\cmidrule(lr){2-3} \cmidrule(lr){4-5} \cmidrule(lr){6-7}
& \textbf{Rate (\%)} & \textbf{N/T} & \textbf{Rate (\%)} & \textbf{N/T} & \textbf{Rate (\%)} & \textbf{N/T} \\
\midrule
SQL Injection & 36.6 & 20/55 & 36.6 & 20/55 & \textbf{74.49} & \textbf{41/55} \\
Cross-Site Scripting (XSS) & 53.98 & 103/191 & \textbf{62.35} & \textbf{119/191} & \textbf{97.93} & \textbf{187/191} \\
Command Injection \& RCE & 34.08 & 31/91 & \textbf{38.38} & \textbf{35/91} & \textbf{90.11} & \textbf{82/91} \\
SSTI & 47.5 & 19/40 & \textbf{50} & \textbf{20/40} & \textbf{97.5} & \textbf{39/40} \\
IDOR \& Access Control & 32.37 & 21/65 & \textbf{57.05} & \textbf{37/65} & \textbf{100} & \textbf{65/65} \\
Authentication Failures & 36.57 & 42/115 & \textbf{50.56} & \textbf{58/115} & \textbf{99.13} & \textbf{114/115} \\
SSRF & 34.85 & 7/20 & \textbf{60.25} & \textbf{12/20} & \textbf{100} & \textbf{20/20} \\
Path Traversal \& LFI & 42.55 & 14/33 & \textbf{54.52} & \textbf{18/33} & \textbf{96.97} & \textbf{32/33} \\
XXE & 20 & 1/5 & \textbf{40} & \textbf{2/5} & \textbf{100} & \textbf{5/5} \\
Cryptographic Failures & 25 & 1/4 & 25 & 1/4 & \textbf{50} & \textbf{2/4} \\
\midrule
Overall & 43.34 & 216/499 & \textbf{52.97} & \textbf{264/499} & \textbf{93.99} & \textbf{469/499} \\
\bottomrule
\end{tabular}
\par\smallskip
\begin{flushleft}
\footnotesize{\textit{Note:} N/T denotes the Number of completed subtasks (N) over the Total subtasks (T).}
\end{flushleft}
\end{table*}

Overall, the results demonstrate consistent performance advantages of Red-MIRROR over both the baseline and current SOTA approaches. The next research question investigates whether fine-tuned mid-scale open-source models can approach comparable performance levels.

\subsubsection{RQ2: Potential of Fine-tuned Mid-scale Open-source LLMs}

To address RQ2, we evaluate the performance of the open-source mid-scale model Qwen2.5-14B (both base and fine-tuned variants) against the large-scale proprietary model DeepSeek-V3.2. Evaluation follows the metrics outlined in Section~\ref{subsec:evaluation_metrics}, focusing on the number of solved challenges and subtask completion rates across vulnerability categories. Due to slower inference speed in the self-hosted setting, Qwen2.5-14B variants are allowed a 30-minute time limit per challenge, while DeepSeek-V3.2 is constrained to 15 minutes.

Results are summarized in \textbf{Table~\ref{tab:rq2-solved}} (solved challenges) and \textbf{Table~\ref{tab:rq2-subtask}} (subtask completion rates).

The base Qwen2.5-14B model exhibits low performance, largely due to insufficient domain-specific cybersecurity knowledge. Fine-tuning yields a clear improvement, increasing solved challenges from 2\% to 12\% on XBOW and overall subtask completion from 43.34\% to 52.97\%. This confirms the effectiveness of domain-specific fine-tuning. Nevertheless, a substantial performance gap persists relative to DeepSeek-V3.2 (86\% solved challenges and 93.99\% subtasks overall), indicating that mid-scale open-source models remain limited in handling complex, multi-step penetration testing tasks even after fine-tuning.

\begin{table*}[!t]
\centering
\caption{Ablation results on XBOW and Vulhub.}
\label{tab:rq3-solved}
\begin{tabular}{llcccc}
\toprule
\textbf{Benchmark} & \textbf{Configuration} & \textbf{Solved} & \textbf{Total} & \textbf{Rate (\%)} & \textbf{Avg TTE} \\
\midrule
\multirow{4}{*}{XBOW}
  & w/o SRMM \& w/o Dual-phase reflection & 22 & 50 & 44 & 21.77 \\
  & w/o Dual-phase reflection             & 32 & 50 & 64 & 17.94 \\
  & w/o SRMM                              & 31 & 50 & 62 & 18.42 \\
  & Full (SRMM + Dual-phase reflection)   & \textbf{43} & \textbf{50} & \textbf{86} & \textbf{13.19} \\
\midrule
\multirow{4}{*}{Vulhub}
  & w/o SRMM \& w/o Dual-phase reflection & 1  & 8  & 12.5 & 7 \\
  & w/o Dual-phase reflection             & 2  & 8  & 25 & 8 \\
  & w/o SRMM                              & 2  & 8  & 25 & 7 \\
  & Full (SRMM + Dual-phase reflection)   & \textbf{4}  & \textbf{8}  & \textbf{50} & \textbf{8} \\
\bottomrule
\end{tabular}
\end{table*}

\subsubsection{RQ3: Ablation Study on SRMM and Dual-phase Reflection}

\subsubsection*{Ablation Results: Effectiveness and Exploitation Efficiency}

To quantify the individual and synergistic contributions of the SRMM and Dual-phase reflection components, we conducted an ablation study using four configurations of the Red-MIRROR system. As described in Sections~\ref{subsec:srmm} and \ref{subsec:reflection}, SRMM maintains long-term contextual coherence across agents, while Dual-phase reflection enables self-evaluation and payload adjustment prior to execution. Each ablation isolates the impact of one or both modules.

Results are reported in \textbf{Table~\ref{tab:rq3-solved}} (overall ablation results on XBOW and Vulhub) and \textbf{Table~\ref{tab:rq3-vuln-ablation}} (vulnerability-wise ablation results across configurations).

\begin{table*}[!t]
\centering
\caption{Vulnerability-wise ablation results across Red-MIRROR configurations.}
\label{tab:rq3-vuln-ablation}

\setlength{\tabcolsep}{5pt}

\begin{tabular}{llm{3cm}m{3cm}cc}
\toprule
\textbf{Vulnerability} & \textbf{Metric}
& \textbf{w/o SRMM \& Dual-phase Reflection}
& \textbf{w/o Dual-phase Reflection}
& \textbf{w/o SRMM}
& \textbf{Full} \\
\midrule

\multirow{3}{*}{SQL Injection}
& SR (\%)  & 52.93 & 60.25 & 60.25 & \textbf{74.49} \\
& Solved   & 0     & 0     & 0     & \textbf{3} \\
& Avg TTE  & 0  & 0  & 0  & \textbf{14.67} \\
\midrule

\multirow{3}{*}{XSS}
& SR (\%)  & 86.41 & 93.21 & 93.21 & \textbf{97.93} \\
& Solved   & 11    & 13    & 13    & \textbf{15} \\
& Avg TTE  & 35.55 & 30.23 & 30.69 & \textbf{18.40} \\
\midrule

\multirow{3}{*}{Command Injection \& RCE}
& SR (\%)  & 68.23 & 85.73 & 80.23 & \textbf{90.11} \\
& Solved   & 2     & 6     & 5     & \textbf{9} \\
& Avg TTE  & 7  & 10.17 & 8.4  & \textbf{9.11} \\
\midrule

\multirow{3}{*}{SSTI}
& SR (\%)  & 59.95 & 72.45 & 72.45 & \textbf{97.5} \\
& Solved   & 2     & 3     & 3     & \textbf{4} \\
& Avg TTE  & 13 & 12.33 & 13.33 & \textbf{12} \\
\midrule

\multirow{3}{*}{IDOR \& Access Control}
& SR (\%)  & 81.43 & 90.74 & 90.74 & \textbf{100} \\
& Solved   & 3     & 4     & 4     & \textbf{7} \\
& Avg TTE  & 8.33  & 9  & 9.5  & \textbf{11.29} \\
\midrule

\multirow{3}{*}{Authentication Failures}
& SR (\%)  & 66.95 & 80 & 80 & \textbf{99.13} \\
& Solved   & 4     & 7     & 7     & \textbf{13} \\
& Avg TTE  & 6.5  & 8.43  & 9.14  & \textbf{11.15} \\
\midrule

\multirow{3}{*}{SSRF}
& SR (\%)  & 75.25 & 100 & 100 & \textbf{100} \\
& Solved   & 2     & 3      & 3      & \textbf{3} \\
& Avg TTE  & 5  & 7.33   & 7.67   & \textbf{7.33} \\
\midrule

\multirow{3}{*}{Path Traversal \& LFI}
& SR (\%)  & 87.94 & 93.94 & 93.94 & \textbf{96.97} \\
& Solved   & 3     & 4     & 4     & \textbf{4} \\
& Avg TTE  & 7.67  & 9  & 9 & \textbf{8.75} \\
\midrule

\multirow{3}{*}{XXE}
& SR (\%)  & 60 & 100 & 100 & \textbf{100} \\
& Solved   & 0     & 1      & 1      & \textbf{1} \\
& Avg TTE  & 0  & 16  & 17  & \textbf{13} \\
\midrule

\multirow{3}{*}{Cryptographic Failures}
& SR (\%)  & 25 & 25 & 25 & \textbf{50} \\
& Solved   & 0     & 0     & 0     & \textbf{0} \\
& Avg TTE  & 0  & 0  & 0 & \textbf{0} \\
\midrule

\multirow{3}{*}{\textbf{Overall}}
& SR (\%)  & 75 & 86 & 85 & \textbf{93.99} \\
& Solved   & 23    & 34    & 33    & \textbf{47} \\
& Avg TTE  & 21.13 & 17.35 & 17.73 & \textbf{12.74} \\
\bottomrule
\end{tabular}
\end{table*}

The ablation results reveal the following:

\begin{itemize}
    \item Removing both components yields a baseline subtask completion rate of 75\%.
    \item Enabling SRMM alone (w/o Dual-phase reflection) increases overall subtask completion to 86\% (+11\%), with particularly strong gains in logic-based and stateful vulnerabilities (IDOR, Authentication).
    \item Enabling Dual-phase reflection alone (w/o SRMM) achieves 85\% (+10\%), showing clear benefit in observable injection families through better payload refinement and error reduction.
\item The full configuration combining both modules reaches the highest performance (93.99\% subtask completion), demonstrating synergistic effects.
\end{itemize}

Average token cost per challenge increases progressively: \$0.10 (no SRMM/Dual-phase reflection), \$0.12 (no Dual-phase reflection), \$0.15 (no SRMM), and \$0.20 (full), compared to \$0.05 for VulnBot. The performance gains justify the additional cost in most practical settings.

Beyond subtask completion rate, the time to exploit (TTE) metric provides additional insight into exploitation efficiency across configurations.

At the benchmark level, TTE exhibits clear separation on XBOW, where many challenges enforce input filtering and require iterative payload refinement. In this setting, the full configuration reduces average TTE from 21.77 (w/o SRMM \& w/o dual-phase reflection) to 13.19, indicating substantially shorter exploitation trajectories in terms of agent steps.

By contrast, Vulhub largely reproduces CVE scenarios without strict input validation or adaptive filtering mechanisms. Exploitation paths are typically short and deterministic, resulting in similar average TTE values across configurations (7--8 steps). The limited variation suggests that, in the absence of defensive constraints, the contribution of additional reasoning mechanisms primarily affects success rate rather than exploitation length.

A finer-grained inspection at the vulnerability level further reinforces this pattern, particularly for injection-based categories. The clearest effect appears in XSS:
\begin{itemize}
    \item For XSS, average TTE decreases from 35.55 (no modules) to 18.40 (full configuration), nearly halving the number of agent steps required among solved instances.
    \item Although TTE does not decrease in these cases, improvements in SQL injection and command injection are primarily reflected in the increased number of solved challenges. Therefore, TTE should not be interpreted in isolation but rather considered jointly with the number of solved instances.
    \item In contrast, vulnerabilities such as SSRF or path traversal exhibit relatively small TTE variation, reflecting more direct exploitation paths once the correct primitive is identified.
\end{itemize}

\begin{table*}[!b]
\centering
\caption{Ablation analysis under different input filtering mechanisms.}
\label{tab:rq3-filter-detailed}
\setlength{\tabcolsep}{3pt}
\scalebox{0.95}{
\begin{tabular}{lcccccccccccc}
\toprule
\multirow{2}{*}{\textbf{Filter Type}} 
& \multicolumn{3}{c}{\textbf{w/o SRMM \& Dual Phase}} 
& \multicolumn{3}{c}{\textbf{w/o Dual Phase Reflection}} 
& \multicolumn{3}{c}{\textbf{w/o SRMM}} 
& \multicolumn{3}{c}{\textbf{Full}} \\
\cmidrule(lr){2-4} \cmidrule(lr){5-7} \cmidrule(lr){8-10} \cmidrule(lr){11-13}
& TTE & Solved & SR(\%) 
& TTE & Solved & SR(\%) 
& TTE & Solved & SR(\%) 
& TTE & Solved & SR(\%) \\
\midrule
Type 1 
& 11.45 & 11 & 71.34 
& 10.68 & 19 & 84.83 
& 10.61 & 18 & 82.78 
& \textbf{10.32} & \textbf{28} & \textbf{95.89} \\

Type 2 
& 31.33 & 9 & 78.97 
& 27.9 & 10 & 83.82 
& 28.4 & 10 & 83.82 
& \textbf{17} & \textbf{12} & \textbf{89.21} \\

Type 3 
& \textbf{7} & 1 & 87.62 
& 10 & \textbf{2} & \textbf{100}
& 10 & \textbf{2} & \textbf{100}
& 9.5 & \textbf{2} & \textbf{100} \\

Type 4 
& 35.5 & 2 & 78.79 
& 29.33 & \textbf{3} & \textbf{100}
& 30 & \textbf{3} & \textbf{100}
& \textbf{19.33} & \textbf{3} & \textbf{100} \\

Type 5 
& -- & 0 & 66.95
& -- & 0 & 86.14 
& -- & 0 & 86.14
& \textbf{14.5} & \textbf{2} & \textbf{100} \\

\midrule
ALL 
& 21.13 & 23 & 75
& 17.35 & 34 & 86 
& 17.73 & 33 & 85
& \textbf{12.74} & \textbf{47} & \textbf{93.99} \\
\bottomrule
\end{tabular}
}
\end{table*}

Overall, the full configuration increases the total number of solved instances (47) while maintaining the lowest average TTE (12.74), demonstrating improvements in both effectiveness and exploitation efficiency.

Importantly, the pronounced TTE separation observed in XBOW and injection-heavy categories suggests that the efficiency gains are closely related to the presence of defensive input filtering mechanisms. This observation motivates the dedicated analysis presented in Section~\ref{sec:rq3-robust}, where challenges are explicitly grouped according to their filtering characteristics (such as pattern-based sanitization, encoding-based escaping, or absence of filtering), and TTE is compared across these groups. By isolating filtering as an experimental factor, we more directly assess whether integrating SRMM and dual-phase reflection enhances pentesting capability under defensive pressure rather than merely improving performance in unconstrained scenarios.

\subsubsection*{Robustness Analysis under Defensive Input Filtering}
\label{sec:rq3-robust}

To directly assess robustness under defensive pressure, we stratify all challenges according to their input filtering mechanisms and compare configuration performance within each group. This analysis isolates filtering strength as the primary experimental variable and evaluates how the relative superiority of different Red-MIRROR configurations evolves under diffenrent defensive constraints.

Specifically, each challenge is categorized into one of the following five filtering types:

\begin{enumerate}
\item \textit{Type 1 – No Filtering.}
No explicit sanitization or validation is applied to user input.
\item \textit{Type 2 – Blocked list Filtering.}  
Specific dangerous substrings are blocked via pattern matching (like keyword-based filtering).

\item \textit{Type 3 – Canonicalized Blocked list Filtering.}  
Inputs are normalized (such as URL-decoded or case-folded) before Blocked list checks, preventing trivial encoding-based bypasses.

\item \textit{Type 4 – Allowlist Filtering.}  
Only predefined safe tokens or formats are permitted, significantly constraining payload construction.

\item \textit{Type 5 – Replacement-Based Sanitization.}  
Potentially dangerous patterns are dynamically removed or rewritten, creating iterative feedback effects across attempts.

\end{enumerate}

For each type, we compare subtask completion rate and TTE across configurations. This setup allows us to determine whether performance gaps—particularly the dominance of configurations incorporating SRMM and/or Dual-phase reflection—expand as filtering strictness increases. If superiority becomes more pronounced at higher filtering levels, it provides direct evidence that the architectural components primarily enhance robustness under constrained input regimes rather than merely improving aggregate performance.

\textbf{Table~\ref{tab:rq3-filter-detailed}} reports the detailed breakdown of Subtask Rate (SR) and average TTE across filtering categories. TTE is computed only over successful exploits (TTE $>$ 0); ``--'' indicates that no successful exploit was observed for that configuration within the corresponding filtering type.

When aggregating results across all filtering categories, the Full configuration achieves the strongest overall performance. It reduces average TTE from 21.13 (w/o SRMM \& Dual Phase Reflection) to 12.74, corresponding to a 39.7\% decrease, while improving subtask completion rate from 75\% to 93.99\% (+18.99 percentage points). Removing either Shared Recurrent Memory Mechanism (SRMM) or Dual Phase Reflection individually still yields improvements over the fully ablated variant, but neither matches the Full configuration. This indicates that both components contribute independently, yet their combination delivers the largest gain.

A clearer separation appears as defensive strength increases. Under Type 2 (Blocked list filtering), the Full configuration reduces TTE from 31.33 (w/o both components) to 17, a 45.7\% reduction. A similar effect is observed under Type 4 (Allowlist filtering), where TTE decreases from 35.5 to 19.33 (45.5\% reduction). In contrast, under Type 1 (No filtering), the reduction is comparatively small (approximately 10\%), suggesting that the architectural advantages become substantially more pronounced once explicit constraints are imposed.

Under Type 3 (Canonicalized Blocked list filtering), all configurations except the fully ablated variant reach 100\% subtask rate. TTE differences narrow, indicating that once trivial encoding-based bypasses are eliminated, success depends more on reasoning correctness than on iterative adaptation speed. In this regime, the benefit of adding one component is primarily reflected in reliability rather than dramatic efficiency gains.

Type 5 (Replacement-based sanitization) represents the strongest defensive setting. The fully ablated configuration and both single-component variants fail to produce any successful exploit, whereas the Full configuration maintains a 100\% subtask rate with an average TTE of 14.5. This outcome highlights a synergistic interaction: Shared Memory alone or Dual Phase Reflection alone is insufficient when payloads are dynamically rewritten across attempts; sustained success requires their integration.

Overall, the relative advantage of the Full configuration increases with filtering strictness. Improvements are modest in unconstrained environments, substantial under Blocked list and Allowlist defenses, and decisive under replacement-based sanitization. This progression supports the conclusion that SRMM and Dual Phase Reflection primarily enhance robustness under constrained and adversarial input regimes rather than merely improving average-case performance.

\subsection{Case Study-based Analysis} \label{subsec:casestudy}
This subsection analyzes selected case studies to qualitatively assess how different Red-MIRROR configurations influence agent behavior, as well as to expose representative failure patterns under sparse-feedback exploitation scenarios.

\subsubsection{Ablation Case Study}
A representative example of successful exploitation is provided by challenge \texttt{XBEN-041-24}, which involves an arbitrary file upload vulnerability leading to remote code execution (RCE). To highlight the impact of different configurations, we compare the attack progression under two settings: (1) Red-MIRROR without Shared Recurrent Memory Mechanism (SRMM) and without Reflection, and (2) Red-MIRROR with full options (including SRMM and Reflection).

In the configuration without SRMM and Reflection, the agent identifies the upload form and successfully uploads a test file (\texttt{test.jpg}), confirming that the file is stored at \texttt{/upload/test.jpg} and that the directory is publicly accessible. However, when attempting to upload a PHP webshell, the agent lacks a systematic mechanism to verify execution or to refine its strategy. After several unsuccessful attempts to access the uploaded file, the process terminates without achieving RCE.

In contrast, the full-option configuration (Red-MIRROR with SRMM and Reflection) produces a coherent and goal-directed exploitation trajectory. Rather than enumerating low-level request details, we summarize the agent behavior at the decision level to highlight the contribution of each module.

\begin{itemize}
    \item Reconnaissance. The agent verifies that uploaded files are stored under a web-accessible \texttt{/upload/} directory.
    \item Payload synthesis. Using SRMM to retain the inferred upload path and URL structure, the agent generates a PHP webshell compatible with the target environment.
    \item Deployment. The webshell is successfully uploaded without server-side filtering or filename modification.
    \item Execution verification. Dual-phase reflection-driven probing confirms command execution by adaptively appending query parameters to the uploaded file.
    \item Objective completion. After an initial failure to read the flag file directly, the Dual-phase reflection module revises the strategy and retrieves the flag via environment variable inspection.
\end{itemize}

To highlight the behavioral differences introduced by SRMM and Reflection, a condensed interaction trace is shown in \textbf{Fig.~\ref{box:fileupload_trace}}.

\noindent
\begin{tcolorbox}[
    float,
    colback=gray!10,
    colframe=black,
    boxrule=0.4pt,
    sharp corners,
    title={File Upload Exploitation (XBEN-041-24)},
    fonttitle=\bfseries,
    width=\linewidth
]
\captionof{figure}{End-to-end exploitation trace demonstrating arbitrary file upload leading to RCE.}
\label{box:fileupload_trace}

\textbf{Configuration:} Red-MIRROR with SRMM and Dual-phase reflection enabled

\medskip
\textbf{Vulnerability:} Arbitrary File Upload \\
\textbf{Impact:} Remote Code Execution (RCE)

\medskip
\textbf{Condensed Interaction Trace:}
\begin{itemize}
    \item \texttt{POST /upload} $\rightarrow$ \texttt{test.jpg}
    \item \texttt{POST /upload} $\rightarrow$ \texttt{shell.php}
    \item \texttt{GET /upload/shell.php?cmd=id}
    \item \texttt{GET /upload/shell.php?cmd=echo \$FLAG}
\end{itemize}

\medskip
\textbf{Observed Behavior:}
SRMM preserves contextual knowledge of the upload path across steps, while the Dual-phase reflection module adaptively revises probing strategies after failed attempts, enabling reliable end-to-end exploitation.

\end{tcolorbox}

This case illustrates how SRMM enables persistent contextual memory across attack steps, while Dual-phase reflection facilitates adaptive hypothesis testing and strategy revision, jointly leading to successful end-to-end exploitation.

\subsubsection{Representative Failure Cases}
\label{sec:failure}

Despite the improvements enabled by the full-option configuration, several challenges remain unsolved even under optimal settings. These failure cases are not specific weaknesses of Red-MIRROR’s architecture, but rather reflect fundamental limitations of LLM-based reasoning under sparse-feedback conditions. These failures consistently arise in scenarios characterized by sparse feedback, framework-specific exploitation requirements, or tasks demanding specialized analytical tooling. The following representative cases illustrate distinct yet recurring failure modes observed across the benchmark.

\paragraph{Case 1: Blind Injection}\mbox{}\\
Blind inference remains a major bottleneck under sparse-feedback conditions, even when the underlying vulnerability is correctly identified.

In challenge \texttt{XBEN-029-24}, the intended attack chain is as follows:
\begin{itemize}
    \item Reconnaissance: Identify the login form endpoint.
    \item Exploitation:
    \begin{enumerate}
        \item Confirm time-based blind SQL injection.
        \item Infer database name.
        \item Extract username(s) from user tables.
        \item Extract corresponding password(s).
        \item Authenticate and access restricted functionality (such as file upload).
        \item Achieve RCE and retrieve the flag.
    \end{enumerate}
\end{itemize}

In practice, the agent completes only the initial stages of this pipeline:
\begin{itemize}
    \item Reconnaissance. The login endpoint and parameters (\texttt{username}, \texttt{password}) are correctly identified, with no evident CAPTCHA or strict rate limiting.
    \item Blind SQL injection detection. Time-based payloads produce consistent latency differences under true conditions, and the Dual-phase reflection module successfully confirms the vulnerability.
    \item Database name inference. Character-by-character extraction using \texttt{SUBSTRING} and \texttt{SLEEP} succeeds, but incurs high request cost and is sensitive to timing instability.
    \item Username extraction (failure). Extraction from user tables fails due to unknown string lengths, record counts, and character sets. The required number of requests grows prohibitively, while network jitter introduces noise that destabilizes inference, causing the process to stall.
\end{itemize}

This failure arises from the intrinsic limitations of purely time-based blind SQL injection (and similarly blind OS command injection), which provides only a binary timing signal. The resulting low information gain per query transforms exploitation into a high-cost search over a large hypothesis space, where autonomous reasoning and reflection mechanisms do not scale effectively.

\paragraph{Case 2: Framework-Specific Exploitation Failure}\mbox{}\\
Certain vulnerabilities require precise, framework-dependent payload placement that cannot be inferred from vulnerability class alone.

In a benchmark based on \texttt{CVE-2022-22963} (Spring Cloud Function), successful exploitation depends on injecting a malicious expression through the HTTP header \texttt{spring.cloud.function.routing-expression}, rather than through the request body or standard parameters. While the agent correctly verifies that the vulnerable endpoint is reachable and processes benign requests as expected, it fails during exploitation by placing the payload in the request body. Consequently, the request is handled normally without triggering code execution.

This case highlights a limitation in handling framework-specific exploit semantics. Technologies with highly specialized invocation mechanisms require exact protocol-level knowledge, and incorrect payload placement results in silent exploitation failure despite correct vulnerability identification.

\paragraph{Case 3: Cryptography-Based Challenges}\mbox{}\\
A third category of failure occurs in challenges involving cryptographic weaknesses.

Such tasks typically require mathematical analysis, algorithmic reasoning, and the construction of custom attack scripts (such as padding oracle exploitation, weak RSA key recovery, or multi-layer encoding reversal). The current Red-MIRROR system does not integrate specialized cryptanalysis or symbolic computation tools. As a result, while the agent may correctly classify a challenge as cryptographic in nature, it lacks the capability to execute the required analytical or computational steps, leading to systematic failure across cryptography-oriented benchmarks.

\paragraph{Summary of failure patterns}
Across all cases, a consistent pattern emerges: Red-MIRROR performs robustly when exploitation steps provide observable, content-rich feedback or follow well-documented patterns. In contrast, performance degrades sharply in scenarios involving sparse or binary feedback (like blind inference), framework-specific payload semantics, or domains requiring specialized analytical tooling such as cryptography. These failure modes delineate the practical boundaries of the current system and motivate targeted extensions, including blind-specific oracle modeling, deeper framework-aware knowledge integration, and hybrid tool-assisted reasoning for cryptographic analysis.

These observations, together with the quantitative results from RQ1--RQ3, are further synthesized and discussed in the Section \ref{sec:discussion}, with a focus on cross-step behaviors, system limitations, and broader implications.

%% file: sections/5-discussion.tex
\section{Discussion}
\label{sec:discussion}

This section synthesizes the experimental findings and provides a cross-cutting analysis of Red-MIRROR’s behavior across different attack stages. Rather than focusing solely on vulnerability categories, we analyze systematic success and failure patterns at the level of individual steps within multi-stage attack chains, followed by a discussion of key limitations and their implications for autonomous penetration testing systems.

\subsection{Cross-cutting Analysis of Attack Steps} \label{subsec:attack_steps}

Results from RQ1--RQ3 consistently indicate that Red-MIRROR’s effectiveness is primarily determined by the characteristics of individual steps (subtasks) within an attack chain, rather than by the vulnerability class alone. Analyzing successful and failed steps across the entire benchmark (50 XBOW challenges and 8 Vulhub CVEs) reveals several recurring patterns.

\paragraph{Observable and Patterned Steps}
Steps with clear, observable feedback and well-established exploitation patterns achieve consistently high success rates. These include reconnaissance steps (such as endpoint discovery, login form identification, default credential detection), stateless injection vulnerabilities with explicit leakage (reflected XSS, error-based or UNION-based SQL injection, SSTI with built-in output channels), and short- to medium-length stateful steps such as IDOR parameter manipulation and authentication privilege escalation.

In these cases, Dual-phase reflection effectively refines payloads based on informative responses (error messages, reflected content, or data dumps), while SRMM maintains session state and tokens across requests. As a result, these steps account for the majority of successful subtasks, reaching an average completion rate of 93.99\% under the Full Option configuration, with particularly strong performance in XSS, IDOR, and authentication failure scenarios.

\paragraph{Blind and Knowledge-intensive Steps}
In contrast, Red-MIRROR exhibits notable weaknesses on steps that require blind inference or specialized domain knowledge. The most prominent failure modes arise from blind SQL injection and blind command injection, where inference relies on timing or boolean side channels. Such steps suffer from low signal-to-noise ratios, as network jitter frequently leads to false positives or negatives. In these settings, Dual-phase reflection becomes largely ineffective due to the absence of semantic output, while SRMM can only store binary delay observations rather than meaningful contextual information. Consequently, brute-force extraction procedures become unstable and often terminate prematurely.

Additional failures occur in framework-specific exploits that require precise payload placement, such as vulnerabilities that mandate injection into non-standard headers rather than request bodies. When such details are absent from the retrieval corpus, agents frequently target incorrect injection points, leading to immediate failure despite successful reconnaissance.

\paragraph{Component-level Effects}
The ablation study further clarifies the complementary roles of SRMM and Dual-phase reflection. SRMM contributes most strongly to stateful steps that require consistent session management, while Dual-phase reflection substantially improves performance for injection-based steps with observable outputs. However, both mechanisms degrade significantly in blind or low-signal settings, limiting their combined effectiveness. This explains why Red-MIRROR performs best in attack chains dominated by feedback-rich, observable steps.

\subsection{Failure Modes and Limitations} \label{subsec:limitations}

While the results demonstrate Red-MIRROR’s effectiveness in structured and observable penetration testing scenarios, several limitations constrain its broader applicability.

\paragraph{Blind and Low-signal Vulnerabilities}
Blind vulnerabilities represent the most challenging class for Red-MIRROR. Time-based and boolean-based attacks rely on indirect side channels and offer only low-bandwidth, binary feedback signals, making inference highly sensitive to network jitter and environmental noise. As a result, numerous requests are often required to achieve statistically meaningful confidence, while minor timing fluctuations can lead to false positives or negatives. In such settings, Dual-phase reflection becomes largely ineffective due to the absence of semantic output, and SRMM is limited to storing coarse-grained delay observations rather than actionable contextual information. Consequently, brute-force extraction procedures frequently become unstable and terminate prematurely.

This limitation reflects a broader mismatch between language-centric reasoning and low-signal exploitation environments. A promising direction for future iterations of Red-MIRROR is to further decouple high-level reasoning from low-level measurement. Rather than having the language model directly orchestrate repetitive, timing-sensitive requests, the LLM can focus on synthesizing parameterized request templates and inference strategies, while specialized tools handle controlled sampling, timing aggregation, and adaptive stopping logic. Such a separation preserves the agent’s reasoning flexibility while improving robustness and efficiency in low-signal exploitation tasks.

\paragraph{Cryptographic Challenges}
Red-MIRROR also exhibits fundamental limitations when handling cryptography-oriented challenges. Although large language models excel at symbolic reasoning and structural pattern recognition, cryptographic exploitation typically requires precise numerical computation, algebraic manipulation, exhaustive search procedures, and the development of non-trivial exploitation scripts. These tasks cannot be reliably performed through language-based reasoning alone. Consequently, failures in cryptographic challenges stem from the absence of integrated computational and scripting capabilities required to execute concrete cryptanalytic procedures, rather than from incorrect high-level vulnerability identification.

A promising direction for improving Red-MIRROR in cryptography-oriented scenarios is to extend the agent architecture with dedicated cryptanalytic tool integration. In such a design, the LLM would be responsible for identifying cryptographic primitives, attack models, and exploitation strategies, while delegating concrete solving tasks—such as key recovery, oracle exploitation, or decoding pipelines—to specialized external tools (such as cryptographic libraries, constraint solvers, symbolic computation engines, or automated script generation frameworks). This separation of responsibilities would allow Red-MIRROR to retain its strength in strategic reasoning while overcoming current limitations in executing computation-heavy and script-intensive operations. Such a hybrid architecture would enhance scalability, precision, and practical applicability in complex cryptographic challenges.

\paragraph{Framework-specific Exploitation Constraints}
The system also exhibits limitations when handling vulnerabilities that require precise, framework-specific exploitation semantics. In several evaluated cases, successful exploitation depends not only on identifying the correct vulnerability class, but also on placing payloads in highly specific request locations, such as custom HTTP headers, framework-defined routing fields, or non-standard parameters. Although Red-MIRROR can correctly recognize the presence of such vulnerabilities during reconnaissance, exploitation may fail when the required payload placement cannot be reliably inferred from observable behavior alone.

This limitation reflects the difficulty of autonomously reasoning about idiosyncratic framework semantics without explicit exploit context, and highlights a boundary where generic vulnerability reasoning does not suffice for reliable exploitation. Improving robustness in such scenarios requires stronger tool-assisted retrieval mechanisms that surface framework-specific exploit examples, payload placement conventions, and concrete request templates, enabling agents to ground exploitation decisions in framework-aware evidence rather than abstract vulnerability knowledge.

\paragraph{Experimental and Evaluation Constraints}
Several experimental factors constrain the generalizability of the reported results. First, the evaluation benchmarks (XBOW and Vulhub) are limited in scale and contain a relatively small proportion of blind or low-signal challenges, which may bias overall performance toward feedback-rich exploitation scenarios. However, given that a substantial proportion of real-world vulnerabilities—such as Reflected XSS, standard SQL Injection, and RCE—naturally produce observable responses, we argue that the results remain highly representative of practical, high-impact penetration testing engagements. Second, all experiments are conducted in controlled laboratory environments that lack real-world defensive mechanisms such as WAFs, strict rate limiting, or adaptive monitoring, which could substantially alter agent behavior and success rates. Finally, the stochastic nature of LLM inference introduces run-to-run variability, while the subtask-based evaluation relies partially on manual decomposition and judgment, potentially introducing subjectivity in complex attack chains.

\subsection{Implications for Autonomous Penetration Testing Systems} \label{subsec:implications}

These findings suggest several implications for the design of autonomous penetration testing agents. First, reflection and memory mechanisms are highly effective but fundamentally constrained by the availability of informative feedback. Second, memory consistency does not compensate for the absence of computational capabilities required for cryptographic analysis or statistical inference. Finally, LLM-based agents should be viewed as orchestration and reasoning components that must collaborate with specialized tools, rather than as standalone solvers for all vulnerability classes.

\subsection{Ethical Considerations} \label{subsec:ethical}

The development of autonomous penetration testing systems, including the Red-MIRROR framework, necessitates a rigorous examination of ethical implications and dual-use risks inherent to offensive AI. While such systems are designed to strengthen defensive capabilities through systematic vulnerability assessment and automated red teaming, their technical capabilities could be repurposed for malicious cyber operations if deployed irresponsibly. Accordingly, Red-MIRROR is explicitly intended for authorized security research and controlled defensive validation environments. To promote responsible use, future iterations should incorporate formal access control mechanisms to ensure that only authenticated and authorized personnel can operate the system. In addition, the system incorporates verifiable activity auditing and cryptographic watermarking of generated payloads to ensure traceability and accountability of AI-driven actions. Constraining knowledge retrieval to trusted and curated repositories further reduces the risk of unintended acquisition of harmful or unverified materials. By embedding these safeguards as foundational design principles rather than auxiliary features, Red-MIRROR advances a paradigm of autonomous security testing that is both technically rigorous and ethically grounded.

\subsection{Threats to Validity}
\label{subsec:threats_validity}

We identify several threats to the validity of our experimental results, categorized into internal, external, and construct validity.

\paragraph{Internal Validity}
Internal validity concerns factors that might influence the causal relationship between the proposed framework and the observed performance improvements.
\begin{itemize}
    \item \textit{Stochasticity of LLMs}: The non-deterministic nature of Large Language Models introduces variability in decision-making and payload generation. Although we set temperature parameters to low values to enforce determinism, run-to-run variations may still occur. To mitigate this, we employed majority voting in critical decision phases; however, complete reproducibility remains challenging.

    \item \textit{Data Contamination}: Public benchmarks (XBOW and Vulhub) may have been included in the pre-training corpora of the evaluated models, potentially inflating performance. To mitigate this, we emphasize zero-shot reasoning and dynamic flag retrieval, which cannot be solved purely through memorization.
\end{itemize}

\paragraph{External Validity}
External validity relates to the generalizability of our findings to real-world scenarios.
\begin{itemize}
    \item \textit{Benchmark Bias}: The evaluation is biased toward observable vulnerabilities with immediate feedback (e.g., reflected XSS, error-based SQLi, command injection with visible output), while blind or low-observability scenarios remain underrepresented. This may overestimate performance in sparse-feedback environments.

    \item \textit{Benchmark Scope}: The Vulhub benchmark focuses on known critical CVEs with relatively short exploitation chains and well-documented payload patterns. This does not fully capture the uncertainty and ambiguity of zero-day or poorly documented vulnerabilities.

    \item \textit{Real-world Complexity}: The evaluation environments (XBOW and Vulhub) do not fully reflect modern production systems, which often include adaptive Web Application Firewalls (WAFs), intrusion detection systems (IDS), and honeypots.

    \item \textit{Environmental Factors}: Experiments were conducted under stable network conditions. In real-world scenarios, latency, jitter, rate limiting, and service instability may degrade system performance, particularly for timing-based attacks.

    \item \textit{Human Role}: While Red-MIRROR demonstrates strong automation capabilities, it is designed to assist rather than replace human penetration testers, as it lacks deeper contextual reasoning for complex business-logic vulnerabilities.
\end{itemize}

\paragraph{Construct Validity}
Construct validity concerns whether the evaluation measures what it intends to measure.
\begin{itemize}
    \item \textit{Subtask Definition}: The analysis of subtask completion relies on manual decomposition of attack chains, introducing subjectivity in defining task boundaries. We mitigate this by aligning subtasks with standard kill-chain phases.

    \item \textit{Prompt Sensitivity}: The performance of LLM-based agents is sensitive to prompt design. Although prompts are standardized across runs, further optimization may influence results.
\end{itemize}

Despite these limitations, the consistent performance improvements observed across multiple benchmarks and vulnerability categories support the robustness of the proposed SRMM and Dual-phase reflection mechanisms.

%% file: sections/6-conclusion.tex
\section{Conclusion and Future Work} \label{sec:conclusion}
This paper presented Red-MIRROR, a multi-agent framework for automated penetration testing that explicitly addresses two core limitations of existing LLM-based systems: lack of persistent state management and insufficient self-adaptive reasoning. By integrating a Shared Recurrent Memory Mechanism (SRMM) with Dual-phase reflection, Red-MIRROR enables coherent long-horizon attack planning, adaptive payload refinement, and reliable session-level exploitation across complex web vulnerabilities. Extensive evaluation on the XBOW benchmark and real-world Vulhub CVEs demonstrates that Red-MIRROR substantially outperforms a strong baseline under identical model settings, while ablation studies confirm the complementary roles of SRMM and Dual-phase reflection.

While Red-MIRROR remains challenged in sparse-feedback scenarios such as blind injection, these results highlight the promise of structured multi-agent reasoning for practical automated penetration testing and motivate future extensions toward low-signal exploitation settings and human-in-the-loop collaborative frameworks.

\section*{Acknowledgement}

This research is funded by Vietnam National University Ho Chi Minh City (VNU-HCM) under grant number NCM2025-26-01.
